\documentclass[useAMS,usenatbib]{mn2e}

\ifx\pdfoutput\undefined
  \usepackage[dvips]{graphicx}
\else                   
 \usepackage[pdftex]{graphicx}
\fi

\usepackage{xspace}
\usepackage{amssymb}
\usepackage{amsmath}
\usepackage[figuresright]{rotating}

\newcommand{\kms}{\mbox{km\,s$^{-1}$}}
\newcommand{\hicat}{{\sc Hicat}\xspace}
\newcommand{\hipass}{{\sc Hipass}\xspace}

\newcommand{\hi}{{\sc Hi}\xspace}
\newcommand{\miriad}{{\sc miriad}\xspace}
\newcommand{\multifind}{{\sc MultiFind}\xspace}
\newcommand{\tophat}{{\sc TopHat}\xspace}

\newcommand{\mbspect}{{\sc mbspect}\xspace}

\newcommand{\ohi}{\mbox{$\Omega_{\rm HI}$}\xspace}
\newcommand{\os}{\mbox{$\Omega_{\rm S}$}\xspace}

\title[The HIPASS Catalogue -- I]{The HIPASS catalogue -- I. Data presentation}

\author[M.J. Meyer et al.]
{M. J. Meyer,$^{1,2}$\thanks{E-mail: martinm@stsci.edu (MM), mzwaan@eso.org (MZ), rwebster@ph.unimelb.edu.au (RW), lister.staveley-smith@csiro.au (LSS)}
M. A. Zwaan,$^{1,3}$
R. L. Webster,$^{1}$
L. Staveley-Smith,$^{4}$\newauthor 
E. Ryan-Weber,$^{1,4}$
M. J. Drinkwater,$^{5}$
D. G. Barnes,$^{1}$
M. Howlett,$^{6}$\newauthor
V. A. Kilborn,$^{7}$
J. Stevens,$^{1}$
M. Waugh,$^{1}$
M. J. Pierce,$^{6}$
R. Bhathal,$^{8}$\newauthor   
W. J. G. de Blok,$^{9}$
M. J. Disney,$^{9}$
R. D. Ekers,$^{4}$
K. C. Freeman,$^{10}$\newauthor
D. A. Garcia,$^{9}$
B. K. Gibson,$^{6}$
J. Harnett,$^{11}$
P. A. Henning,$^{12}$\newauthor
H. Jerjen,$^{10}$
M. J. Kesteven,$^{4}$
P. M. Knezek,$^{13}$
B. S. Koribalski,$^{4}$ \newauthor
S. Mader,$^{4}$
M. Marquarding,$^{4}$  
R. F. Minchin,$^{9}$
J. O'Brien,$^{10}$ \newauthor
T. Oosterloo,$^{14}$
R. M. Price,$^{12}$ 
M. E. Putman,$^{15}$
S. D. Ryder,$^{16}$\newauthor  
E. M. Sadler,$^{17}$ 
I. M. Stewart,$^{18}$
F. Stootman,$^{8}$     
and A. E. Wright$^{4}$\\
$^{1}$ School of Physics, University of Melbourne, VIC 3010, Australia\\
$^{2}$ Space Telescope Science Institute, 3700 San Martin Drive, Baltimore MD 21218, U.S.A.\\
$^{3}$ European Southern Observatory, Karl-Schwarzschild-Str. 2, 85748 Garching b. M{\"u}nchen, Germany\\
$^{4}$ Australia Telescope National Facility, CSIRO, P.O. Box 76, Epping, NSW 1710, Australia\\
$^{5}$ Department of Physics, University of Queensland, QLD 4072, Australia\\
$^{6}$ Centre for Astrophysics and Supercomputing, Swinburne University of Technology, P.O. Box 218, Hawthorn, VIC 3122 Australia.\\
$^{7}$ Jodrell Bank Observatory, University of Manchester, Macclesfield, Cheshire, SK11 9DL, U.K.  \\
$^{8}$ Department of Physics, University of Western Sydney Macarthur, P.O. Box 555, Campbelltown, NSW~2560, Australia\\
$^{9}$ Department of Physics \& Astronomy, University of Wales, Cardiff, P.O. Box 913, Cardiff CF2 3YB, U.K.\\
$^{10}$ Research School of Astronomy \& Astrophysics, Mount Stromlo Observatory, Cotter Road, Weston, ACT~2611, Australia\\
$^{11}$ University of Technology Sydney, Broadway NSW 2007, Australia\\
$^{12}$ Institute for Astrophysics, University of New Mexico, 800 Yale Blvd, NE, Albuquerque, NM~87131, USA.\\
$^{13}$ WIYN, Inc. 950 North Cherry Avenue, Tucson, AZ, U.S.A.\\
$^{14}$ ASTRON, P.O. Box 2, 7990 AA Dwingeloo, The Netherlands\\
$^{15}$ CASA, University of Colorado, Boulder, CO 80309-0389, U.S.A.\\
$^{16}$ Anglo-Australian Observatory, P.O. Box 296, Epping, NSW~1710, Australia\\
$^{17}$ School of Physics, University of Sydney,  NSW~2006, Australia\\
$^{18}$ Department of Physics \& Astronomy,  University of Leicester, Leicester LE1 7RH, U.K.}

\begin{document}

\date{Accepted ...
      Received ...}

\pagerange{\pageref{firstpage}--\pageref{lastpage}}
\pubyear{0000}

\maketitle

\label{firstpage}

\begin{abstract}
The \hi Parkes All-Sky Survey (\hipass) Catalogue forms the largest
uniform catalogue of \hi sources compiled to date, with 4,315 sources
identified purely by their \hi content.  The catalogue data comprise
the southern region $\delta < +2\degr$ of \hipass, the first blind \hi
survey to cover the entire southern sky.  RMS noise for this survey is
13 mJy beam$^{-1}$ and the velocity range is $-$1,280 to 12,700~\kms.
Data search, verification and parametrization methods are discussed
along with a description of measured quantities.  Full catalogue data
are made available to the astronomical community including positions,
velocities, velocity widths, integrated fluxes and peak flux
densities.  Also available are on-sky moment maps, position-velocity
moment maps and spectra of catalogue sources.  A number of local
large-scale features are observed in the space distribution of sources
including the Super-Galactic plane and the Local Void.  Notably,
large-scale structure is seen at low Galactic latitudes, a region
normally obscured at optical wavelengths.
\end{abstract}
 

     \begin{table*}
     \caption{Parameters of existing blind \hi surveys. $\delta v$ is
     the velocity resolution. $N^{lim}_{HI}$ is the limiting \hi
     column density for gas filling the beam.  Column density limits
     are $5\sigma$ and calculated with $\Delta V = 100$\kms, scaling
     RMS appropriately according to the velocity resolution.  SCC, BGC
     and \hicat all use \hipass data.  The SCC sample searches $\delta
     < -62\degr$ and the BGC searches $\delta < 0\degr$ and $S_{p} >$
     116 mJy.  The references cited are as follows:
     $^1$\citet{shostak1977},
     $^2$\citet{krumm1984},$^3$\citet{kerr1987}, \citet{henning1992},
     $^4$\citet{sorar1994},\citet{zwaan1997},\citet{spitzak1998},
     $^6$\citet{rosenberg2000}, $^7$\citet{henning2000},
     $^8$\citet{kilborn2002}, \citet{braun2003}, $^{10}$\citet{lang2003},
     $^{11}$\citet{koribalski2003}, $^{12}$ this paper.}
     \label{tab:blind_surveys}
     \begin{minipage}{17.5cm}
     \begin{center}
     \small
     \begin{tabular*}{17.5cm}{p{19mm}p{12mm}p{15mm}p{23mm}p{10mm}p{10mm}p{14mm}p{25mm}p{8mm}}
     \hline
     Name                                       & Area                                & Beam Size & Velocity                                          & $\delta v$  & RMS      & $N^{lim}_{HI}$ & Telescope & Sources  \\
                                                & (deg$^2$)                           & (arcmin)  & (\kms)                                            & (\kms)    & (mJy beam$^{-1}$)  & (cm$^{-2}$)                &           &          \\
     \hline	
     Shostak$^1$                                & {\raggedright 85,\\70,\\11}         & 10.8      & {\raggedright $\sim$$-$800~to~$\sim$2,835}        & 11        & {\raggedright 32,\\44,\\18}    & 1.6$\times 10^{19}$, 2.3$\times10^{19}$, 9.2$\times 10^{18}$ & {\raggedright 91m Green Bank} & 1$^{a}$ \\ 
     {\raggedright Krumm~\&\\ Brosch$^2$}       & {\raggedright $\sim$35,\\$\sim$44}  & 10.8      & {\raggedright 6,300~to~9,600,\\ 5,300~to~8,500}   & $\sim$45  & 22          & 2.2$\times 10^{19}$  & {\raggedright 91m Green Bank} &  0\\
     Henning$^3$                                & 7,204 pointings                     & 10.8      & {\raggedright $-$400~to~7,500}                    & 22        & -           & -                    & {\raggedright 91m Green Bank} & 37 \\
     AHISS$^4$                                  & 65                                  & 3.3       & {\raggedright $<$7,400}                           & 16        & 0.75        & 4.3$\times 10^{18}$  & {\raggedright 305m Arecibo} & 66\\
     Slice Survey$^5$                           & 55                                  & 3.3       & {\raggedright 100 to 8,340}                       & 16        & 1.7         & 9.6$\times 10^{18}$  & {\raggedright 305m Arecibo} & 75\\
     ADBS$^{6}$                                 & 430                                 & 3.3       & {\raggedright $<$7,980}                           & 34        & 3.5         & 2.8$\times 10^{19}$  & {\raggedright 305m Arecibo} & 265\\
     HIZSS$^{7}$                                & 1,840                               & 15.5      & {\raggedright $-$1,200 to 12,700}                 & 27        & 15          & 5.0$\times 10^{18}$  & {\raggedright 64m Parkes}   & 110\\
     SCC$^{8}$$^{\S}$                           & $\sim 2400$                         & 15.5      & {\raggedright $-$1,280~to~12,700}                 & 18        & 13          & 4.0$\times 10^{18}$  & {\raggedright 64m Parkes} & 536\\
     \raggedright{WRST Wide Field Survey$^{9}$} & 1,800                               & 49        & {\raggedright $-$1,000~to~6,500}                  & 17        & 18          & 4.8$\times 10^{17}$  & {\raggedright Westerbork Array} & 155  \\
     HIJASS$^{10}$$^{b}$                       & 1,115                               & 12        & {\raggedright $-$3,500~to~10,000}                 & 18        & 16          & 7.3$\times 10^{18}$  & {\raggedright 76m Jodrell Bank} & 222  \\
     BGC$^{11}$$^{\S}$                          & 20,626                              & 15.5      & {\raggedright $-$1,200~to~8,000}                  & 18        & 13          & 1.7$\times 10^{19}$$^{d}$ & {\raggedright 64m Parkes} & 1,000\\
     \hicat$^{12}$$^{\S}$                       & 21,346                              & 15.5      & {\raggedright 300~to~12,700}                      & 18$^{c}$  & 13          & 4.0$\times 10^{18}$  & {\raggedright 64m Parkes} & 4,315\\
     \hline
     \hline
     \multicolumn{9}{l}{$^{a}$Not certainly extragalactic.}\\
     \multicolumn{9}{l}{$^{b}$Survey in progress.}\\
     \multicolumn{9}{l}{$^{c}$Data velocity resolution. Parametrization was carried out with additional Hanning smoothing, giving a velocity resolution of 26.4 \kms.}\\
     \multicolumn{9}{l}{$^{d}$Based on catalogue selection limits rather than 5$\sigma$ survey limit.}\\
     \end{tabular*}
     \normalfont
     \end{center}
     \end{minipage}
     \end{table*}

\section{Introduction}

The neutral hydrogen (\hi) content of galaxies provides a unique and
fundamental perspective into the nature and evolution of galaxies,
being intimately linked to both the dark matter component of galaxies
along with their ability to form stars.

Surveys for \hi in galaxies, however, are limited by the relative
weakness of \hi photons.  Despite a cosmological mass density ratio of
\hi to luminous matter of only 1:10,\footnote{\ohi $=
(3.8\pm0.6)\times 10^{-4} h_{75}^{-1}$ \citep{zwaan2003}
cf. \os$=(39\pm5.7)\times 10^{-4} h_{75}^{-1}$ \citep[Salpeter initial
mass function,][]{cole2001}} the improbability of the \hi hyperfine
transition and weakness of photons emitted mean that
the received power ratio favours optical photons by many orders of
magnitude.  In combination with observational limitations, this has
traditionally necessitated either blind surveys of relatively small
volumes (eg. \citealt{shostak1977}, \citealt{krumm1984},
\citealt{kerr1987}, \citealt{henning1992}, \citealt{sorar1994},
\citealt{zwaan1997}, \citealt{spitzak1998}, \citealt{rosenberg2000}),
or targeted surveys across larger regions through the use of existing,
typically optical, catalogues (eg. \citealt{fisher1981},
\citealt{mathewson1992}, \citealt{giovanelli1997},
\citealt{haynes1999}).  Parameters for existing blind \hi surveys are
summarized in Table~\ref{tab:blind_surveys}.

With the advent of Multibeam receivers on the Parkes and Jodrell Bank
radio telescopes, along with the forthcoming ALFA instrument at
Arecibo, large untargeted surveys are now practicable for the first
time.  In this work we discuss results from the \hi Parkes All-Sky
Survey (\hipass), a blind survey of the entire sky $\delta<+25\degr$.
Various extragalactic catalogue subsets from this survey have already
been completed including: \citet[][$\delta<-62$]{kilborn2002};
\citet[][$\delta < 0\degr$ and $S_{\rm p} > 116$ mJy]{koribalski2003}
and companion analysis papers \citet[][\hi mass function]{zwaan2003}
and \citet[][newly catalogued galaxies]{ryan2002}; \citet[][Fornax
region]{waugh2002}; and \citet[][Centaurus A group]{banks1999}.
Presented here are results from the southern region of the survey
($\delta<+2\degr$); these data were searched for extragalactic sources
to full survey depth.

A complete database of \hi-selected galaxies is important for a number
of important astrophysical studies.  Amongst these is a measurement of
the \hi cosmic mass density, \ohi.  At higher redshifts, \ohi can be
measured through damped Ly$\alpha$ systems, however at $z$=0, the only
way to accurately measure \ohi is through large-scale surveys for \hi
emission.  Galaxies under-represented in optical surveys such as low
surface brightness galaxies, those with elusive optical counterparts
and Malin 1 type systems \citep{bothun1987} are also best detected via
blind \hi surveys.  Gaining an accurate census of such objects is
important for constraining models of galaxy formation, although not
contributing a significant fraction of the overall gas mass density
\citep{zwaan2003}.

Other important applications for a large uniform catalogue of \hi
sources include the measurement of large-scale structure, 2-point
correlation functions, the Tully-Fisher relation, the role of hydrogen
gas in the evolution of galaxy groups and the study of environments
around \hi-rich galaxies compared to those of optically-selected
galaxies.

In Section~\ref{sec:hipass_survey} we discuss the observational
methods, data processing and basic properties of \hipass.  Methods
used to compile the source database are described in detail in
Section~\ref{sec:catalogue}, including the identification,
verification and parametrization of sources, confused- and
extended-source processing, the effects of radio-frequency
interference (RFI) and recombination lines, and follow-up
observations.  Access to catalogue data is described in
Section~\ref{sec:data}.  Finally, basic property distributions are
presented in Section~\ref{sec:properties} and the large-scale
distribution of galaxies in Section~\ref{sec:lss}.  Reliability,
completeness and accuracy of catalogue parameters are discussed in the
companion paper by \citet[][Paper II]{zwaan2003b}.  The identification
of optical counterparts to \hipass Catalogue galaxies is made in
\citet[][Paper III]{drinkwater2003}.

\section{The HI Parkes All-Sky Survey}
\label{sec:hipass_survey}

\hipass was completed at the Parkes\footnote{The Parkes telescope is
part of the Australia Telescope which is funded by the Commonwealth of
Australia for operation as a National Facility managed by CSIRO} 64m
radio telescope using the Multibeam receiver, an hexagonal array of 13
circular feed-horns installed at the focal plane of the telescope.
Observations and data reduction for this survey are described in
detail in \citet{barnes2001} and \citet{staveley-smith1996}.  A brief
summary is provided below.

\subsection{Observations}

\hipass observations were carried out from February 1997 to March 2000
for the southern portion of the survey ($\delta < +2\degr$) and to
December 2001 for the northern extension ($\delta < +25\degr$).  For
these observations, the telescope was scanned across the sky in
$8^\circ$ strips at a rate of $1^\circ$ per minute, with scans
separated by 7 arcmin in RA, giving a total effective integration time
of 450s beam$^{-1}$.  To obtain spectra, the Multibeam correlator was
used with a 64 MHz bandwidth and 1024 channel configuration, giving a
velocity range of $-$1,280 to 12,700~\kms and a channel separation of
13.2~\kms\, at $z=0$.  Parameters for the survey are summarized in
Table~\ref{tab:hipass}.
       
\subsection{Data Processing}

     \begin{table} 
     \begin{center} 
     \caption{\hipass parameters} \label{tab:hipass} 
     \begin{tabular}{ll} 
     \hline \hline
     Sky Coverage & $\delta <+25\degr$$^{\star}$ \\
     Integration time per beam & 450s \\ 
     Average FWHM & 14.3 arcmin \\
     Gridded FWHM & 15.5 arcmin \\ 
     Pixel size & 4 arcmin \\ 
     Velocity Range & $-$1,280 to 12,700~\kms \\
     Channel separation$^{\dagger}$ & 13.2~\kms \\ 
     Velocity Resolution & 18.0~\kms \\
     Positional accuracy$^{\ddagger}$ & 1.5 arcmin \\
     RMS noise & 13 mJy beam$^{-1}$ \\
     \hline \hline 
     \multicolumn{2}{l}{$^{\star}$ The \hipass Catalogue covers southern regions $\delta <+2\degr$}\\
     \multicolumn{2}{l}{$^{\dagger}$ at $z=0$}\\
     \multicolumn{2}{l}{$^{\ddagger}$ $1\sigma$ accuracy at 99\% completeness flux limit}\\
     \multicolumn{2}{l}{\citep{zwaan2003b}}\\
     \end{tabular}
     \end{center} 
     \end{table}

     \begin{figure}
     \begin{center}
     \includegraphics[trim=0.cm 0cm 0cm 0cm,width=8cm,clip=true]{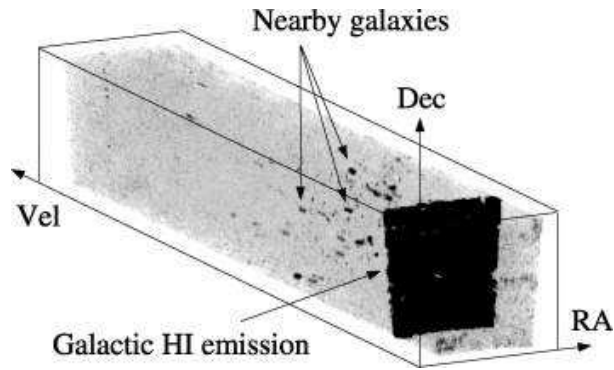}
     \end{center}
     \caption{Orthogonal projection of HIPASS cube 93 showing emission from
     the Milky Way as well as other nearby galaxies.  Velocity range is
     truncated at 10,000 \kms.}
     \label{fig:cube}
     \end{figure}

Initial processing of \hipass data was completed in real-time at the
observatory.  To correct spectra for the standard bandpass effects,
spectra were reduced using the package {\sc livedata}.  This package
also performed the conversion to heliocentric rest frame velocities by
shifting spectra using Fourier techniques.

Bandpass correction is done by dividing the signal target spectrum by
a reference off-source spectrum, this spectrum representing the
underlying spectral shape caused by signal filtering, as well as the
temperature of the receivers, ground and sky.  For the \hipass data,
the reference spectrum for each receiver is obtained by taking signals
immediately before and after a particular integration as the telescope
scans across the sky.  To provide a robust measure of the bandpass,
the value for each channel in the reference spectrum is taken to be
the median rather than the mean of available signals.  However, one
artifact that can be created by this process is the generation of
negative side lobes north and south of particularly bright \hi
sources, since actual \hi emission is included in the bandpass
determination.  Finally, spectra are smoothed with a Tukey filter to
suppress Gibbs ringing, resulting in a velocity resolution of 18~\kms.
 
To create the three-dimensional position-position-velocity cubes that
form the main data product of \hipass, the bandpass-corrected spectra
are gridded together using the package {\sc Gridzilla}.  The spectrum
at each RA-Dec pixel is taken as the median of all data within 6
arcmin of the pixel centre, multiplied by 1.28 to correctly scale
point source flux densities.  This procedure thus corrupts data for
extended sources in the standard \hipass data, with peak flux
densities\footnote{`peak flux density' is hereafter referred to as
`peak flux'} for such sources over-estimated (by 1.28 for an {\em
infinitely} extended source).  Extended sources nevertheless
represent only a small fraction of overall sources ($\sim$2\%, see
Section~\ref{sec:extended}) and the corresponding corrections are
small.  We therefore apply no correction to measured catalogue
parameters for this effect.  The resulting cubes from the gridding
process are $\,8^{\circ} \times 8\,^{\circ}$ with $4' \times 4'$
pixels in the on-sky directions and extend over the full \hipass
velocity range in the third axis (an example cube is shown in
Figure~\ref{fig:cube}).  Overlap regions between cubes vary, but in
general are $\sim1\degr$ in RA and Dec. In total, 388 cubes are
required to cover the southern sky.

Lastly, a correction is applied to the spectra in each \hipass data
cube to minimize baseline distortion caused by the Sun and other
continuum sources.  This is applied using the program {\sc Luther} and
involves the fitting and subtraction of a template distortion
spectrum.

\section{The Catalogue}
\label{sec:catalogue}

The \hipass Catalogue (\hicat) is compiled using a combination of
automatic and interactive processes, with candidate detections first
generated through automated finder scripts and then manually verified.
Detections are finally parametrized using semi-automated routines.
These are discussed below; a summary of the number of detections
remaining at each stage is given in Table~\ref{tab:numbers}.

\subsection{Candidate Generation}

     \begin{table} 
     \begin{center} 
     \caption{Detection numbers at different stages of catalogue compilation} \label{tab:numbers} 
     \begin{tabular}{ll} 
     \hline
     Stage & Number\\
     \hline
     Generated by \multifind & 137,060\\
     Generated by \tophat & 17,232\\
     Examined in first two verification stages & 142,276\\
     Examined in third verification stage & 61,276\\
     Final catalogue & 4,315\\
     \hline \hline 
     \end{tabular}
     \end{center} 
     \end{table}

     \begin{table*}
     \caption{\hipass parameter descriptions. Unless otherwise stated, all velocities are in the heliocentric restframe and $cz$ convention.}
     \label{tab:param_desc}
     \begin{minipage}{17.8cm}
     \begin{tabular*}{17.8cm}{p{15mm}p{17mm}p{14mm}p{115mm}}       
     \hline
     Parameter               & Database Name      & Units         & Description\\
     \hline		        	      	   
     Name                    & {\tt hipass\_name} & $-$           & names are of the form HIPASS JXXXX$\pm$YY[a-z], where XXXX is the unrounded source RA in hrs and min, and YY is the unrounded source declination in degrees.  An additional letter a-z is added if needed to distinguish sources. Names are also made consistent with existing published sources where necessary.\\
     RA                      & {\tt ra}           & hrs           & right ascension (J2000 hexadecimal format)\\
     Dec                     & {\tt          dec} & deg           & declination (J2000 hexadecimal format)\\
     $ v_{\rm 50}^{\rm max}$ & {\tt   vel\_50max} & \kms          & {\raggedright average of velocities at which profile reaches 50 per cent of peak flux density\\ (using width maximization procedure,$^a$ recommended velocity)}\\
     $ v_{\rm 50}^{\rm min}$ & {\tt   vel\_50min} & \kms          & {\raggedright average of velocities at which profile reaches 50 per cent of peak flux density\\ (using width minimization procedure$^a$)}\\
     $ v_{\rm 20}^{\rm max}$ & {\tt   vel\_20max} & \kms          & {\raggedright average of velocities at which profile reaches 20 per cent of peak flux density\\ (using width maximization procedure$^a$)}\\
     $ v_{\rm 20}^{\rm min}$ & {\tt   vel\_20min} & \kms          & {\raggedright average of velocities at which profile reaches 20 per cent of peak flux density\\ (using width minimization procedure$^a$)}\\
     $   v_{\rm mom}$        & {\tt     vel\_mom} & \kms          & flux weighted velocity average between $v_{\rm lo}$ and $v_{\rm hi}$ \\
     $    v_{\rm Sp}$        & {\tt      vel\_Sp} & \kms          & velocity at which profile peak flux occurs\\
     $   v_{\rm gsr}$        & {\tt     vel\_gsr} & \kms          & {\raggedright Galactic standard of rest velocity\\ \citep[converted from heliocentric $v_{\rm mom}$ using][]{devaucouleurs1991}}\\
     $    v_{\rm lg}$        & {\tt      vel\_lg} & \kms          & {\raggedright Local Group standard of rest velocity\\ \citep[converted from heliocentric $v_{\rm mom}$ using][]{karachentsev1996}}\\
     $   v_{\rm cmb}$        & {\tt     vel\_cmb} & \kms          & {\raggedright Cosmic Microwave Background standard of rest velocity\\ \citep[converted from heliocentric $v_{\rm mom}$ using][]{fixsen1996}}\\
     $v_{\rm lo}$            & {\tt      vel\_lo} & \kms          & manually specified minimum profile velocity\\
     $v_{\rm hi}$            & {\tt      vel\_hi} & \kms          & manually specified maximum profile velocity\\
     $v_{\rm speclo}$        & {\tt  vel\_speclo} & \kms          & manually specified minimum velocity for spectral plots and RMS measurement\\
     $v_{\rm spechi}$        & {\tt  vel\_spechi} & \kms          & manually specified maximum velocity for spectral plots and RMS measurement\\
     $v_{\rm mask}$          & {\tt    vel\_mask} & \kms          & manually specified mask regions over which to interpolate baseline fitting and to exclude values from RMS measurement (in velocity pairs)\\
     $ W_{\rm 50}^{\rm max}$ & {\tt width\_50max} & \kms          & {\raggedright difference of velocities at which profile reaches 50 per cent of peak flux density \\(using width maximization procedure$^a$)}\\
     $ W_{\rm 50}^{\rm min}$ & {\tt width\_50min} & \kms          & {\raggedright difference of velocities at which profile reaches 50 per cent of peak flux density \\(using width minimization procedure$^a$)}\\
     $ W_{\rm 20}^{\rm max}$ & {\tt width\_20max} & \kms          & {\raggedright difference of velocities at which profile reaches 20 per cent of peak flux density \\(using width maximization procedure$^a$)}\\
     $ W_{\rm 20}^{\rm min}$ & {\tt width\_20min} & \kms          & {\raggedright difference of velocities at which profile reaches 20 per cent of peak flux density \\(using width minimization procedure$^a$)}\\
     $     S_{\rm p}$        & {\tt           Sp} & Jy            & peak flux density of profile\\
     $   S_{\rm int}$        & {\tt         Sint} & Jy \kms       & integrated flux of source (within region $v_{\rm lo}$, $v_{\rm hi}$ and box size)\\
                RMS          & {\tt          rms} & Jy            & RMS of masked spectrum between $v_{\rm speclo}$ and $v_{\rm spechi}$\\
      RMS$_{\rm clip}$       & {\tt    rms\_clip} & Jy            & clipped RMS of masked spectrum between $v_{\rm speclo}$ and $v_{\rm spechi}$\\
      RMS$_{\rm cube}$       & {\tt    rms\_cube} & Jy            & unsmoothed RMS of cube (median absolute deviation from median multiplied by $\sqrt{\pi/2}$)\\
              cube           & {\tt         cube} & $-$           & cube number\\
           $\sigma$          & {\tt        sigma} & \kms          & standard deviation of Gaussian used in baseline smoothing\\
           box size          & {\tt      boxsize} & arcmin        & box size used for parameter measurements\\
            comment          & {\tt      comment} & $-$           & comment (1=real, 2=have concerns)\\
      follow-up              & {\tt      nb\_flg} & $-$           & narrow band follow-up status flag (1=real)\\
      confused               & {\tt    cfsd\_flg} & $-$           & confused source flag (1=confused)\\
      extended               & {\tt     ext\_flg} & $-$           & extended source flag (1=extended).  For extended sources $S_{\rm int}$ is summed flux within region rather than default point-source weighting.\\
     \hline
     \hline
     \multicolumn{4}{l}{$^a$ Maximization procedure starts at profile velocity limits and searches inward until the required per cent of peak flux}\\
     \multicolumn{4}{l}{is reached.  The minimization procedure starts at
       $v_{\rm S_{p}}$ and searches outward.}
     \end{tabular*}
     \end{minipage}
     \end{table*}

To generate candidate detections, the results of two separate scripts
are combined.  The first of these, a modified version of \multifind
\citep{kilborn2001}, uses a peak flux threshold method, while the
second, \tophat, cross-correlates spectra with tophat profiles of
various scales.  The two methods are used together as neither is able
to recover fully the real sources detected by the other (\multifind
locates $\sim$83\% of final catalogue sources, while \tophat finds
$\sim$90\%).  Both candidate lists are then merged and automatically
cleaned to remove high velocity clouds (HVCs) and suspect detections
at known interference frequencies as detailed below.

\subsubsection{\multifind}

The first of the finder algorithms is based on \miriad
\citep{sault1995} routines.  Each of the velocity planes in a data
cube is searched for detections rising above a specified peak flux
density and a Gaussian fitted to each pixel clump.  For \hicat, the
peak flux threshold is set to four times the calculated noise.  Cube
noise is calculated robustly using the relation:

\begin{equation}
\text{cube noise} = s\times\sqrt{\pi/2},
\end{equation}

\noindent where $s$ is the median absolute deviation from the median.
A median noise estimate is used to provide a method insensitive to
outliers, with the applied conversion equivalent to estimating RMS of
the normally distributed component of cube noise.  Detections are
filtered to include only those detections present in adjacent velocity
planes with a separation $<5$ arcmin.

This detection process is repeated twice, each time Hanning smoothing
the data cube to increase detection signal-to-noise, and cube noise
re-measured.  Finally, detection lists are combined and duplicated
detections removed.

\subsubsection{\tophat}

In the second finder algorithm, each spectrum in the cubes is searched
for emission on a variety of velocity width scales between 1 and 40
channels.  An initial list of detections is created as follows.
First, solar and continuum ripple in the spectral baselines is reduced
by using a moving median filter with a width dependent on the current
search scale.  Spectra are then cross-correlated with a top hat of the
appropriate size, with the convolution weighted by the noise in each
velocity plane.  A feature is detected in a convolved spectrum if it
rises above a threshold proportional to the interquartile range values
in that spectrum.

In the second stage of candidate identification, detections from all
scales are island grouped, listing together all detections that are
either adjacent or overlapping (the majority of features will be
detected on multiple scales).  The bounding box of these groups is
then found, and the original data re-fitted to create the entry in the
final detection list for each group.

\subsubsection{Cleaning}

Results from the automated finders are then merged and cleaned.
First, all candidates with $v_{\rm gsr} < 300$~\kms$\,$ are removed to
avoid Galactic \hi emission and HVCs \citep[HVCs in \hipass are
catalogued separately, see][]{putman2002}.  Additionally, an initial
filter is applied to remove false detections arising from hydrogen
recombination lines and radio frequency interference.  All detections
having velocity widths $< 25.0$~\kms\, in the ranges $2,638 < v_{\rm
hel} < 2,670$~\kms, $4,355 < v_{\rm hel} < 4,397 $~\kms, $4,505 <
v_{\rm hel} < 4,559$~\kms\, and $7,428 < v_{\rm hel} < 7,482$~\kms\,
are removed.  We note that many more recombination line and radio
frequency interference detections are discarded in subsequent
verification procedures than by this simple and conservative filter
(see Sections \ref{sec:interference}~\&~\ref{sec:recombination} for
further discussion).  As no sources in the final catalogue have a
velocity width $< 25$~\kms, these cuts are not expected to have any
effect on the final catalogue completeness.  A declination limit of
$\delta < +2\degr$ is also applied.  Following this process, $\sim
140,000$ candidates remain to be manually checked.

\subsection{Candidate Verification}

     \begin{figure*}
     \begin{center}
     \includegraphics[trim=2.8cm 0cm -1cm 0cm,height=6cm,keepaspectratio=true,angle=270,clip]{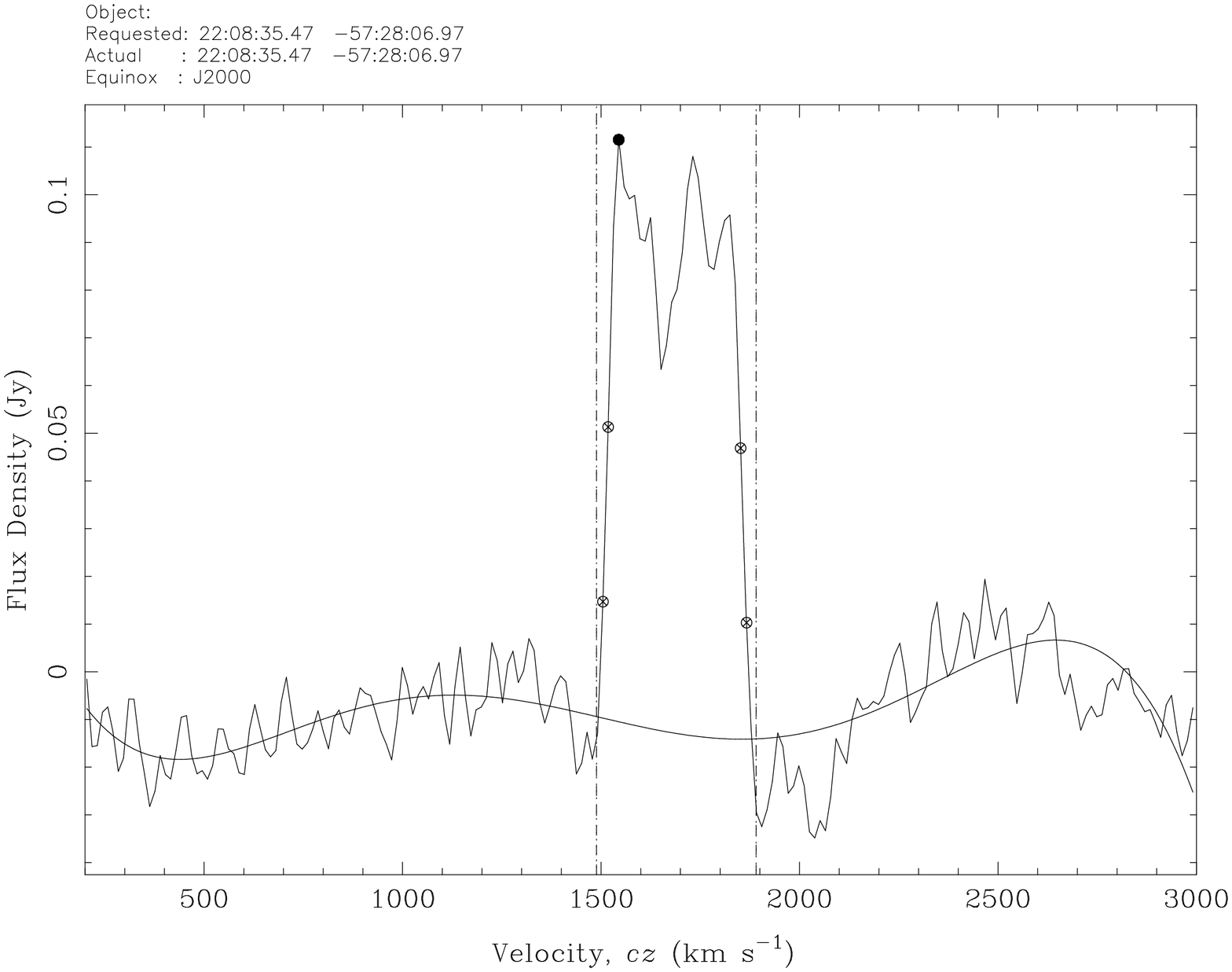}
     \includegraphics[trim=2.8cm 0cm -1cm 0cm,height=6cm,keepaspectratio=true,angle=270,clip]{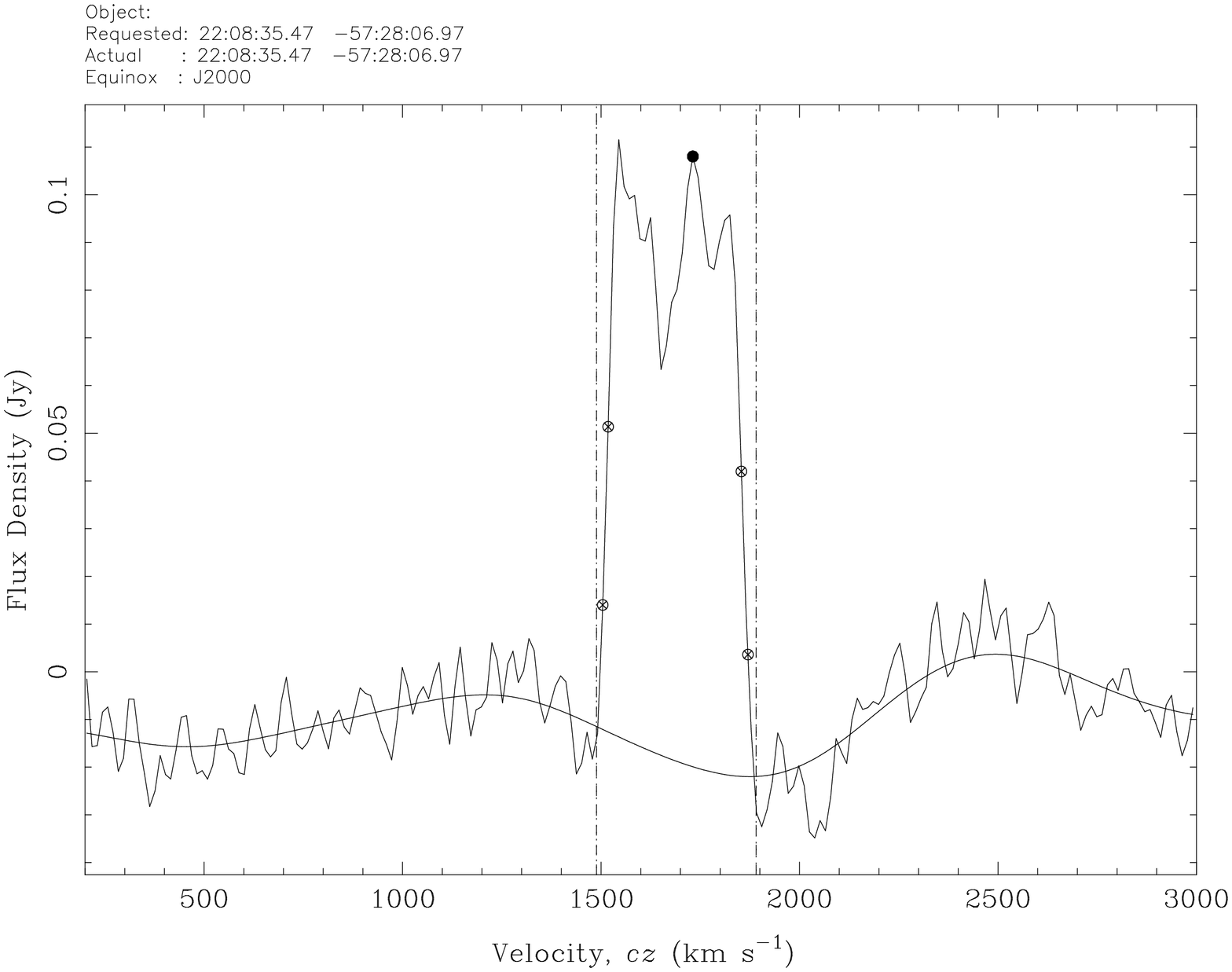}
     \includegraphics[trim=2.8cm 0cm -1cm 0cm,height=6cm,keepaspectratio=true,angle=270,clip]{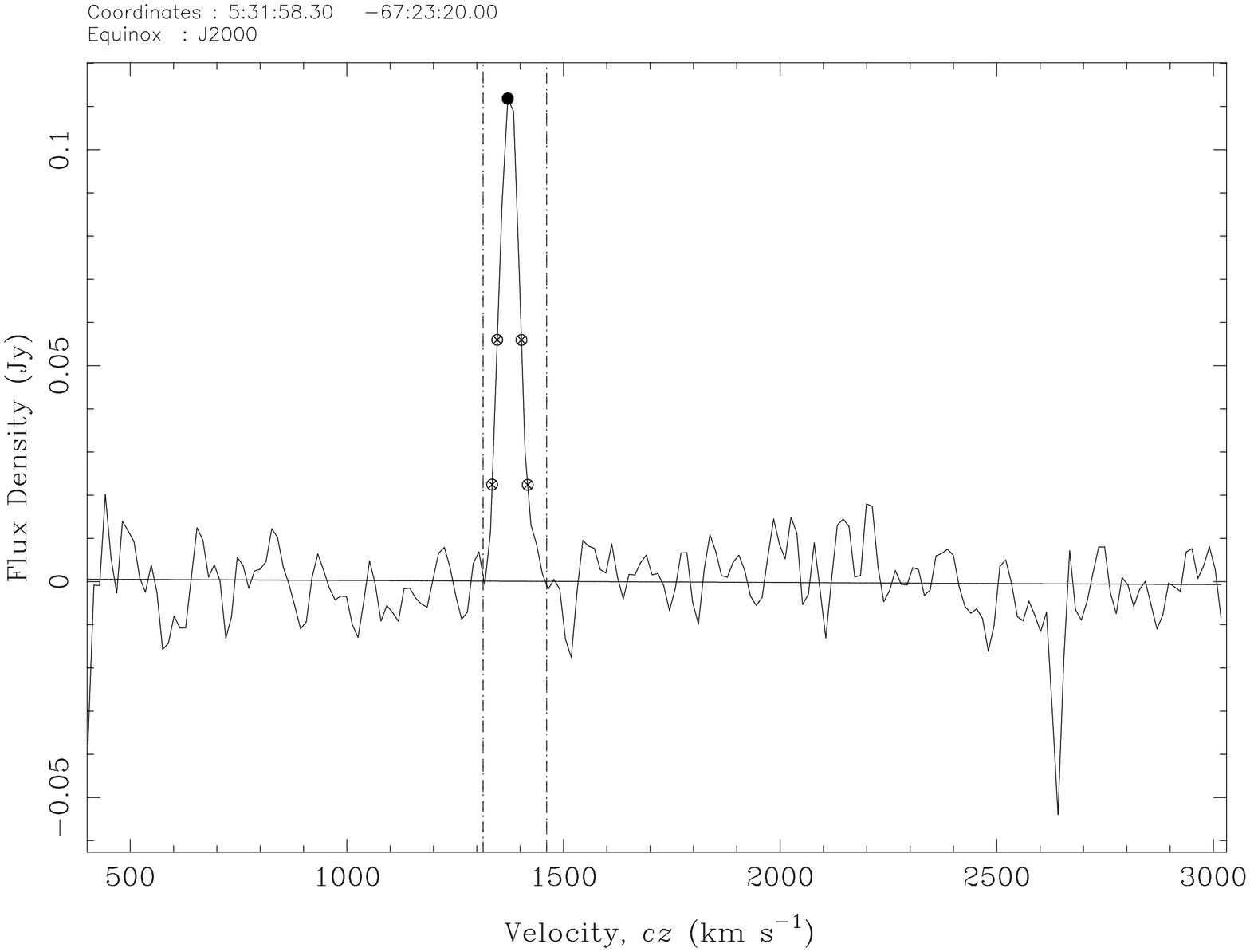}
     \includegraphics[trim=2.8cm 0cm -1cm 0cm,height=6cm,keepaspectratio=true,angle=270,clip]{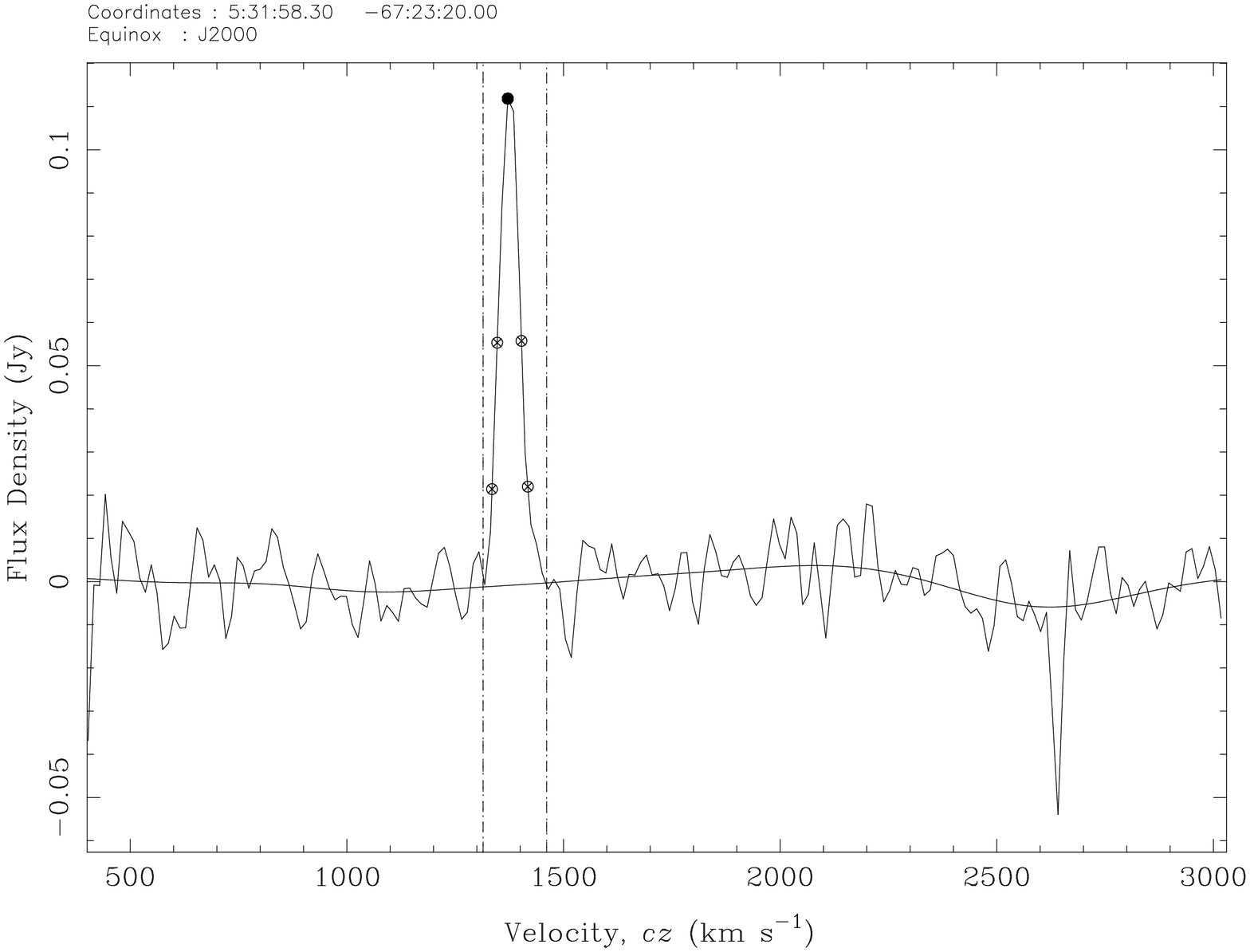}
     \end{center}
     \caption{Examples of baseline fitting using polynomials (left) and Gaussian smoothing (right).  Upper polynomial fit uses a 5th order polynomial and lower fit uses 1st order polynomial.}
     \label{fig:baseline}
     \end{figure*}

     \begin{figure*}
     \begin{center}
     \includegraphics[trim=1cm 0cm 0cm 0cm,height=7cm,keepaspectratio=true,angle=270]{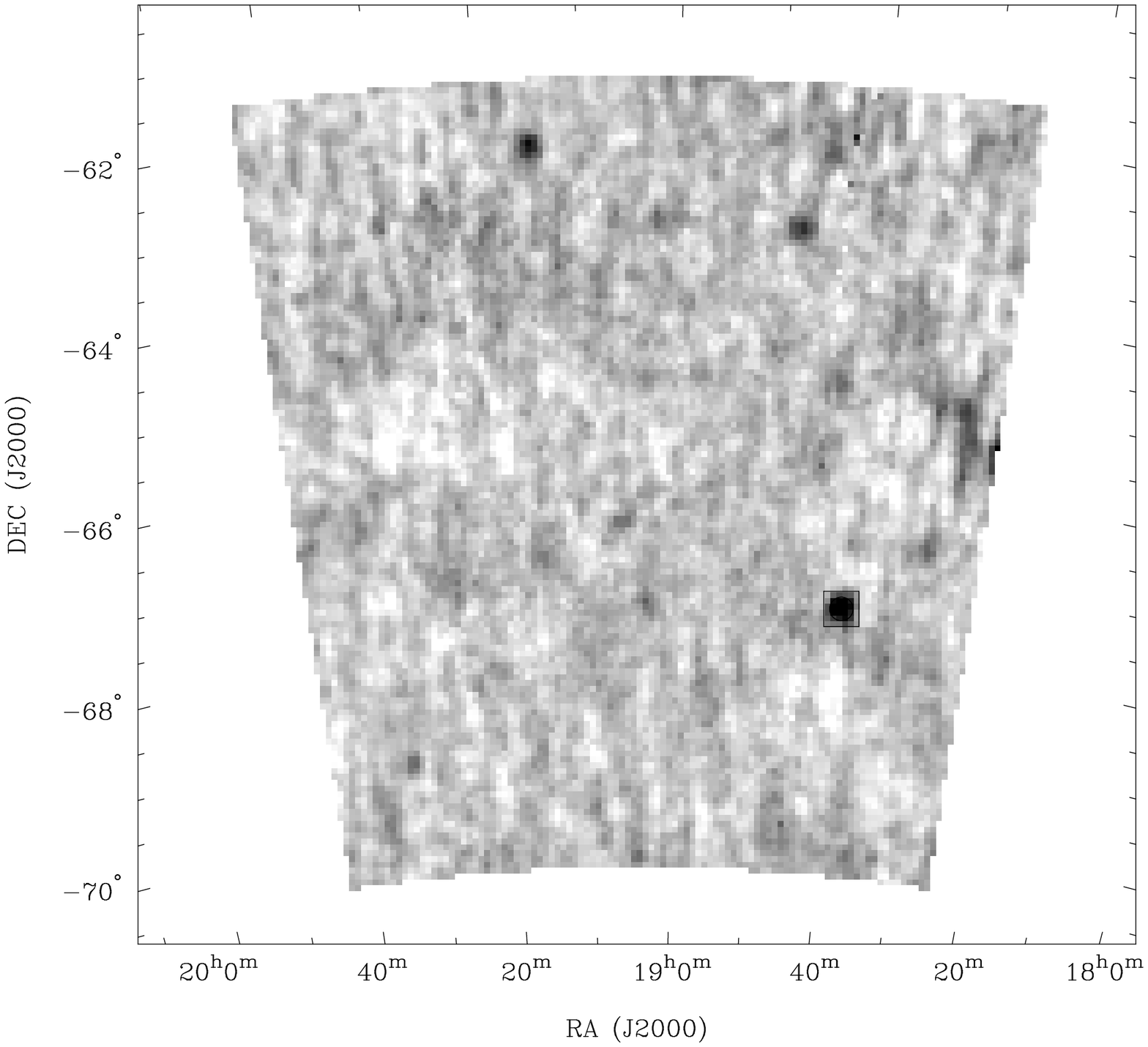}
     \includegraphics[trim=2.8cm 0cm 0cm 0cm,height=8cm,keepaspectratio=true,angle=270,clip]{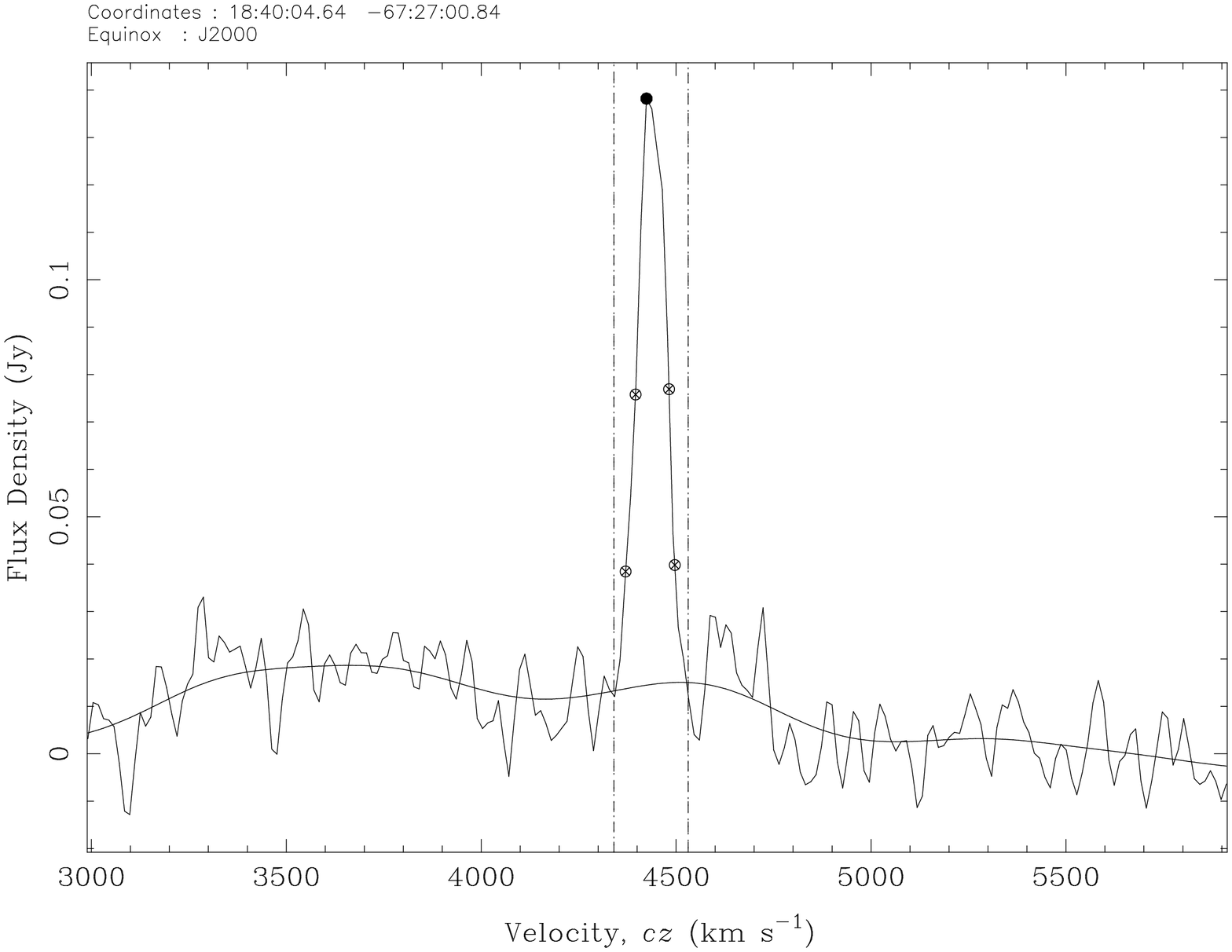}
     \includegraphics[trim=-1cm 0cm 0cm 0cm,height=10cm,keepaspectratio=true,angle=270]{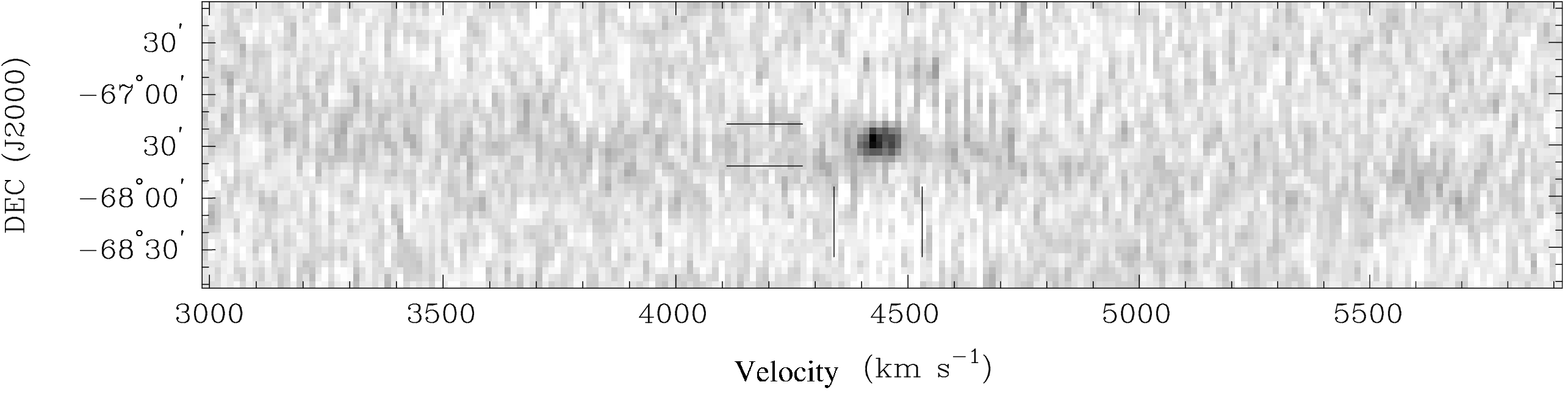}
     \caption{Example plots examined during the final checking and
     parametrization processes: (top-left) RA-Dec moment map showing the
     default $28\times28$ arcmin data box used for position fitting;
     (top-right) spectrum within the $28\times28$ arcmin box, treating the
     detection as a point source and weighting pixels by Parkes' beam
     parameters - vertical lines show manually specified velocity limits
     within which profile parameters are measured and the RA-Dec moment map
     generated; (bottom) Dec-velocity moment map showing 28 arcmin box and
     profile velocity limits.}
     \label{fig:images}
     \end{center}
     \end{figure*}

Verification of candidate sources is a multi-stage process.  First,
two independent checks are made on each candidate, visually inspecting
the full \hipass spectrum summed from a $12\times12$ arcmin box around
the position of the detection. The aim of this process is to quickly
remove obviously spurious detections without rejecting any real
sources.

Next, as a preliminary pass at removing multiple detections of the
same object, candidates are grouped using an angular separation limit
of 5 arcmin and a velocity separation of 100~\kms.  In this process a
detection is added to an existing group if it is within the separation
limits of any of its members.  One detection from each group is then
selected for further examination, giving preference first to those not
rejected in both of first checks, then to the detection with the
highest peak flux and finally, in instances of equal peak flux, to the
detection with the greatest velocity width.  Group size warning limits
are set to 10 arcmin and 200~\kms.  Groups with sizes exceeding these
warning limits are later inspected to ensure no nearby galaxies are
removed (see Section~\ref{sec:confused}).

In the third stage of detection verification, candidates not rejected
by both of the first checks are examined in spectral, position, and
position-velocity space.  Third checks are carried out by one of only
three selected observers to improve consistency.  Positions are fitted
using automatic iterative fitting routines based on original detection
coordinates.  To generate spectra, candidates are treated as
unresolved, with pixels in a $28\times 28$ arcmin box used to
calculate source flux in a given velocity plane via:

\begin{equation}
S(v)=\sum_i\frac{f_i S_i}{f_i^2}
\end{equation}

\noindent where $S_i$ is the flux at a given pixel (and velocity $v$),
and $f_i$ is the value of the Gaussian beam at that pixel, relative to the
previously fitted centre of emission.

While every effort is made to either reject or confirm detections, any
detections unable to be evaluated either way are left in the catalogue
at this stage.  Spurious detections are flagged in the parametrization
stage for follow-up observations.

     \begin{figure}
     \begin{center}
     \includegraphics[trim=1cm 6cm 0cm 0cm,width=8cm,keepaspectratio=true]{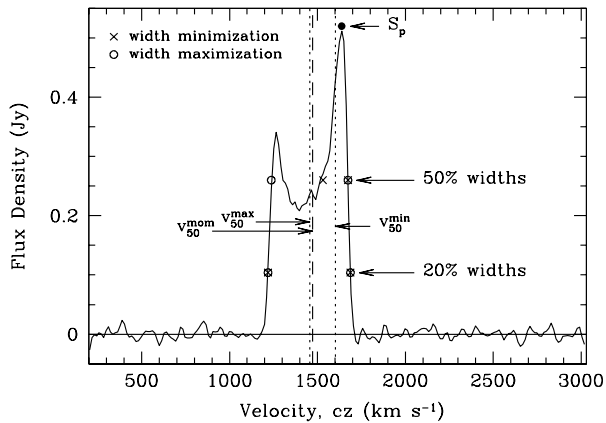}
     \caption{Example spectrum following baseline fitting and removal
     showing the locations of $S_{\rm p}$, $W_{\rm 50}^{\rm max}$, $W_{\rm
     50}^{\rm min}$, $W_{\rm 20}^{\rm max}$, $W_{\rm 20}^{\rm min}$,
     $v_{\rm 50}^{\rm max}$, $v_{\rm 50}^{\rm min}$, $v_{\rm mom}$, $v_{\rm
     lo}$ and $v_{\rm hi}$. The parameters $v_{\rm 20}^{\rm max}$ and
     $v_{\rm 20}^{\rm min}$ are not shown but are defined similarly to
     $v_{\rm 50}^{\rm max}$ and $v_{\rm 50}^{\rm min}$. The offset of
     $v_{\rm mom}$ toward higher velocities from $v_{\rm 50}^{\rm max}$ is
     due to the asymmetric nature of profile.}
     \label{fig:params}
     \end{center}
     \end{figure}

\subsection{Parametrization}

The parametrization of \hicat sources is done interactively using
standard \miriad routines, together with a modified version of
\mbspect that uses Gaussian smoothing for baseline fitting.  All
parametrization is done by M.M. and M.Z.

\subsubsection{Baseline Fitting}

Baseline fitting is carried out by first linearly interpolating the
spectrum over the galaxy profile, as specified with the parameters
$v_{lo}$ and $v_{hi}$ (see Table~\ref{tab:param_desc}).  The average
values of the ten channels on either side of the masked profile are
taken as the endpoints for the interpolation.  The resulting spectrum
is then convolved with a Gaussian of an interactively specified width.
A Gaussian is used to provide the smoothest estimate of the spectral
baseline.  It is important in choosing the width of the Gaussian
kernel to ensure that the scale is large enough to fit smoothly
beneath the galaxy profile without creating spurious small scale
variations.  The Gaussian also needs to be wide enough not to reduce
artificially the noise of the observed spectrum.  At the same time,
the Gaussian should be narrow enough to fit the baseline without
missing the peaks and troughs of the spectrum.  Example spectra
comparing Gaussian smoothing baseline fitting to polynomial baseline
fitting are shown in Figure~\ref{fig:baseline}.

\subsubsection{Parameter Measurement}
  
To parametrize a detection, the position of the candidate is first
refined by fitting an elliptical Gaussian to a $28\times28$ arcmin
moment map around galaxy finder coordinates.  Provided fitting is
successful, the coordinates of the detection are taken as those of the
ellipse and the $28\times28$ arcmin moment map regenerated.  Entering
an interactive loop for parameter measurement, an ellipse is re-fitted
to the new detection location, and the measurement region again moved
to the ellipse position if the fit is successful.  Plots of the
detection are now generated in RA-Dec, Dec-velocity and spectral
spaces, along with initial parameter measurements (see
Figure~\ref{fig:images}).  Spectra are Hanning smoothed for parameter
measurement to improve signal-to-noise, giving a final velocity
resolution of 26.4 \kms.  Spectrum and profile limits, masked regions,
initial fitting position and final position can all be varied using
the interactive windows at this point.  Checks can be made on the
veracity of the detection by changing grey-scales, box size, Hanning
smoothing and the velocity range displayed.  The size of the Gaussian
kernel used in baseline fitting can also be adjusted.  For each
adjustment made, images and moment maps are regenerated, where
appropriate, with fitting redone accordingly.  When a satisfactory fit
is achieved, the detection is given a final comment to confirm or
reject it, with a follow-up flag available if necessary.  By default,
all fluxes are measured using the point source method, treating
sources as non-extended with the central high signal-to-noise pixels
preferentially weighted according to Parkes' beam parameters.  The
treatment of extended sources is discussed in
Section~\ref{sec:extended}. A full list of parameters measured is
given in Table~\ref{tab:param_desc}, along with an example spectrum
illustrating various measured quantities in Figure~\ref{fig:params}.

\subsection{Confused Source Identification and Duplicate Removal}
\label{sec:confused}

To combine the lists from individual cubes, the full set of results is
again island-grouped and cleaned to remove duplicate detections in the
cube overlap regions.  Accurate positions enable much tighter matching
limits, now chosen at 1.5 arcmin and 20~\kms.  Tight limits also avoid
accidentally removing real source pairs.

As the scales of duplicate detections and confused source pairs
overlap at this point, all galaxies with velocity profiles overlapping
and on-sky position within 6$\sigma$ of each other (as defined by
the gridded Parkes beamwidth) are examined manually.  Also included in
this process are any detections previously flagged as confused during
parametrization, as well as galaxies originating from cleaned groups
that exceeded size warning limits in either of the island grouping
stages.  Confused sources flagged in the final catalogue are either
those manually identified at this stage, or detections with velocity
profiles overlapping and source separations $<4\sigma$.

Sources that could reasonably be distinguished (ie. measured fluxes
largely due to the galaxy itself rather than a close neighbour), are
identified and measured separately, whereas those for which this was
not possible are grouped together into a single source and the
detection manually flagged as confused.

\subsection{Extended Sources}
\label{sec:extended}

     \begin{figure}
     \begin{center}
     \includegraphics[trim=0cm 0.5cm 0cm 0cm,width=8cm,keepaspectratio=true]{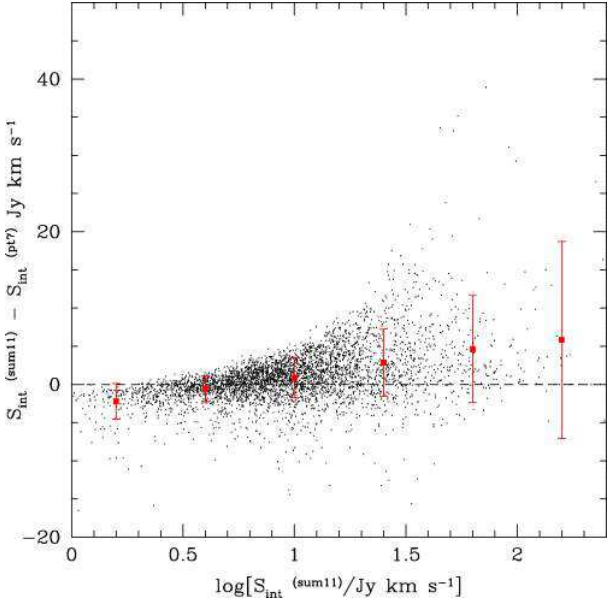}
     \end{center}
     \caption{Difference between summed flux in $44\times44$ arcmin box and
       flux in $28\times28$ arcmin box weighted by Parkes beam parameters
       assuming a point source.  Values increase toward higher integrated
       fluxes as a result of extended sources.}
     \label{fig:flux_extended}
     \end{figure}
     
The minimum expected number of extended sources in \hicat can be
roughly estimated using the relation between \hi mass and \hi diameter
found for an optically selected sample of galaxies
\citep{broeils1997}:

\begin{equation}
\log(M_{HI}) = (1.96 \pm 0.04) \log(D_{HI}) + (6.52 \pm 0.06),
\end{equation}

\noindent where $M_{HI}$ is \hi mass in $M_{\odot}$ and $D_{HI}$ is
the \hi diameter defined at a surface density of 1 $M_\odot pc^{-2}$
in kpc.  Combining this with the relation between \hi mass (in solar
masses), distance (in Mpc) and observed flux $\int S dV$ ($=  S_{\rm
  int}$, in Jy~\kms):

\begin{equation}
M_{HI} = 2.356\times10^5 D^2 \int S dV
\end{equation}

\noindent and making the approximation $1.96\log(D) \thickapprox
\log(D^2)$, this gives a relation for source size that depends solely
on source flux and is independent of distance ($\theta_{HI}$ is the
source diameter in arcminutes):

\begin{equation}
\label{eqn:size}
\int S dV \thickapprox 1.2 \theta_{HI}^2
\end{equation}

The relationship between $\int S dV$ and diameter can be
empirically confirmed by examining the difference between summed
integrated flux in a $44\times44$ arcmin box and the standard
non-extended source integrated flux in a $28\times28$ arcmin box
(pixels weighted by the Parkes beam parameters).  This difference is
plotted as a function of the summed $44\times44$ arcmin integrated
flux in Figure~\ref{fig:flux_extended}.  As expected, the average
difference is close to zero at small integrated fluxes (point sources)
and positive for large integrated fluxes (extended sources), a
positive value indicating that some flux was missed through use of the
point-source method.  The large amount of scatter in this graph is
indicative of the uncertainties in measuring integrated fluxes by
summing the value of all pixels in a given region, compared to the
point source method.

     \begin{figure}
     \begin{center}
     \includegraphics[trim=0cm 5cm 0cm 0cm,width=7.5cm,keepaspectratio=true]{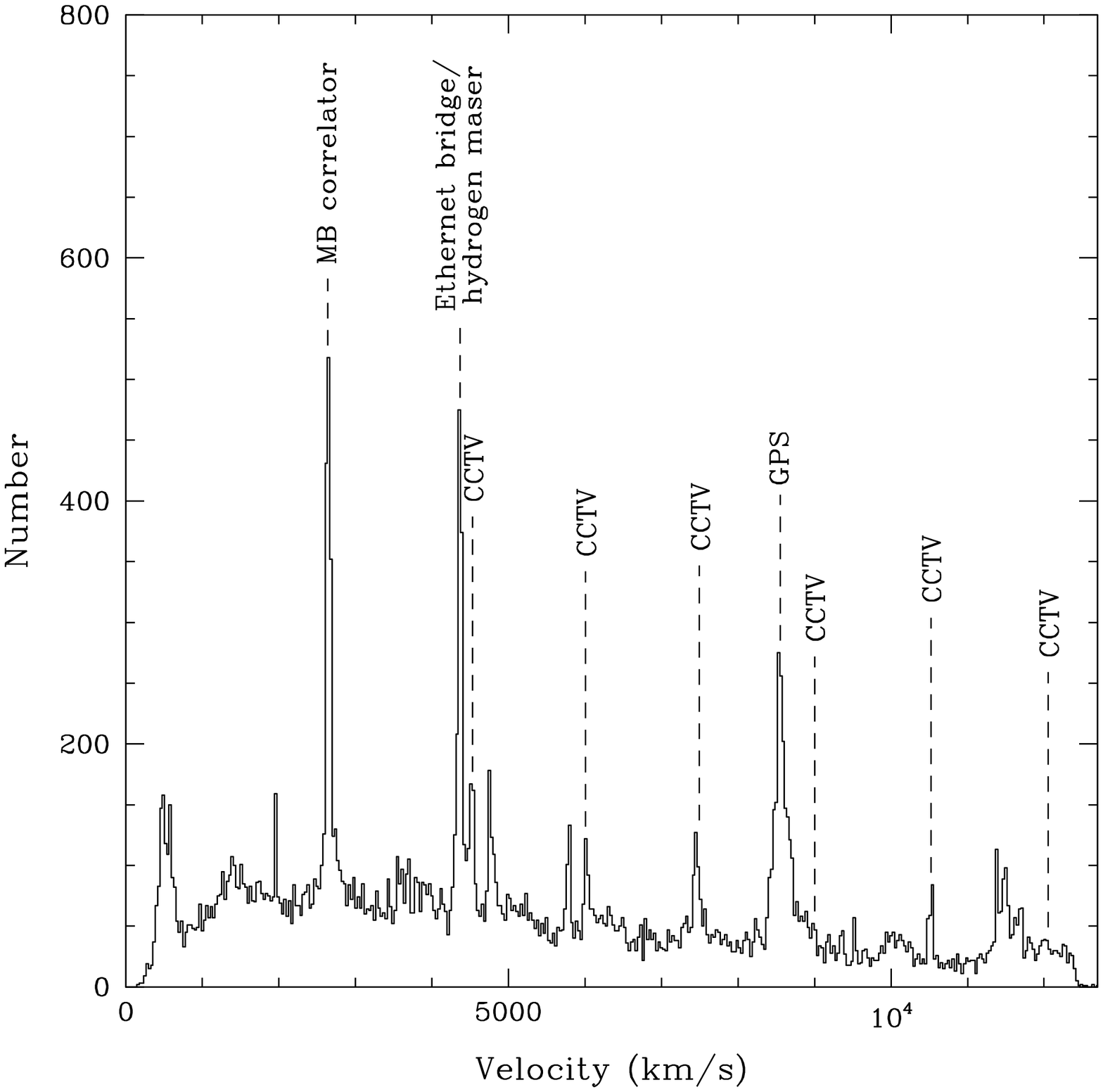}\\
     \includegraphics[trim=0cm 5.5cm 0cm 0cm,width=7.5cm,keepaspectratio=true]{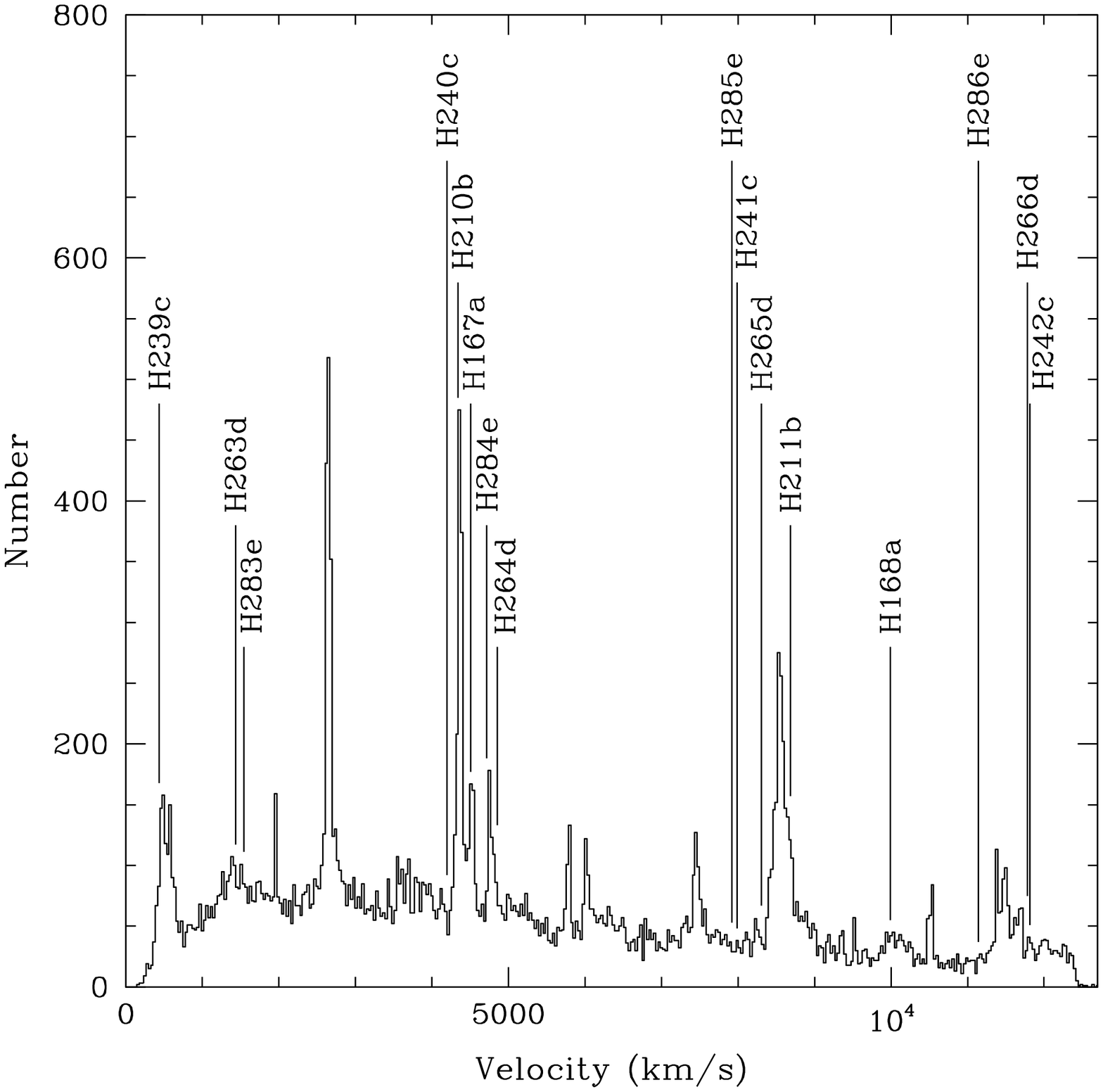}
     \end{center}
     \caption{Velocity distribution of \hipass galaxies after the second
     verification check, with labels corresponding to potential RFI
     frequencies (top) and hydrogen recombination frequencies (bottom).}
     \label{fig:interference}
     \end{figure}

To identify extended sources in \hicat, all sources greater than $7'$
in size are taken as potentially extended, this limit corresponding to
57 Jy~\kms\, from Equation~\ref{eqn:size}.  Assuming a uniform galaxy
flux distribution, 93\% of source flux is retrieved for a $7'$
diameter source if treated as unresolved, and higher for a centrally
biased distribution.  In total, there are 188 candidate sources
brighter than this flux limit, where flux is measured using an
$44\times44$ arcmin `summed' box.  Each source is examined manually,
re-fitted and flagged if necessary.  Where extended sources are
confused or only marginally extended, detections remain fitted as
point sources if this best removes erroneous flux contributions from
neighbouring galaxies or cube noise.  In the final catalogue, 90
sources are flagged as extended.

\subsection{Interference}
\label{sec:interference}

     \begin{figure}
     \begin{center}
     \includegraphics[trim=3.0cm 0cm 1cm 0cm,height=7.5cm,keepaspectratio=true,angle=270,clip]{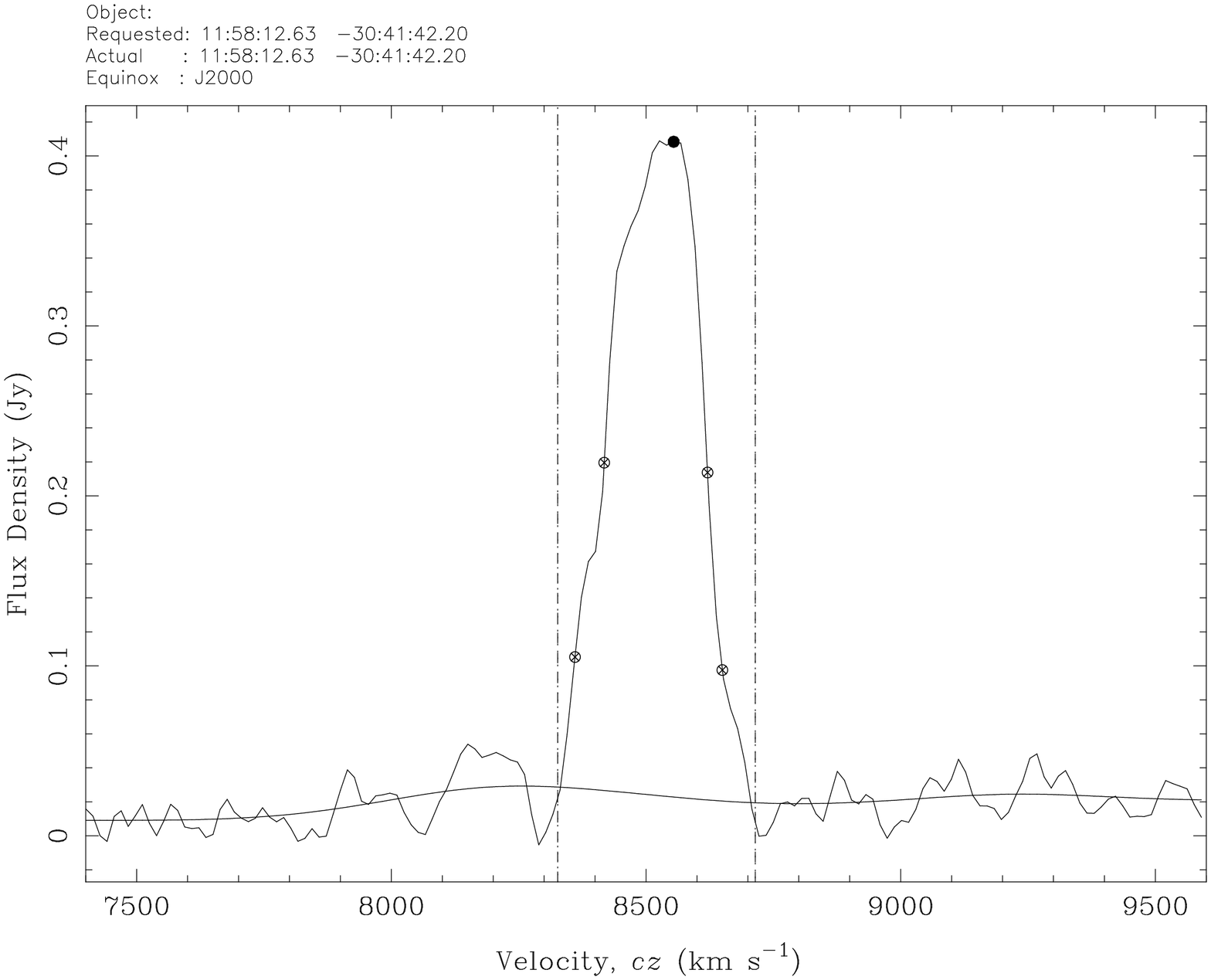}\\
     \includegraphics[trim=6.4cm 1cm 0cm -1cm,height=8.3cm,angle=270]{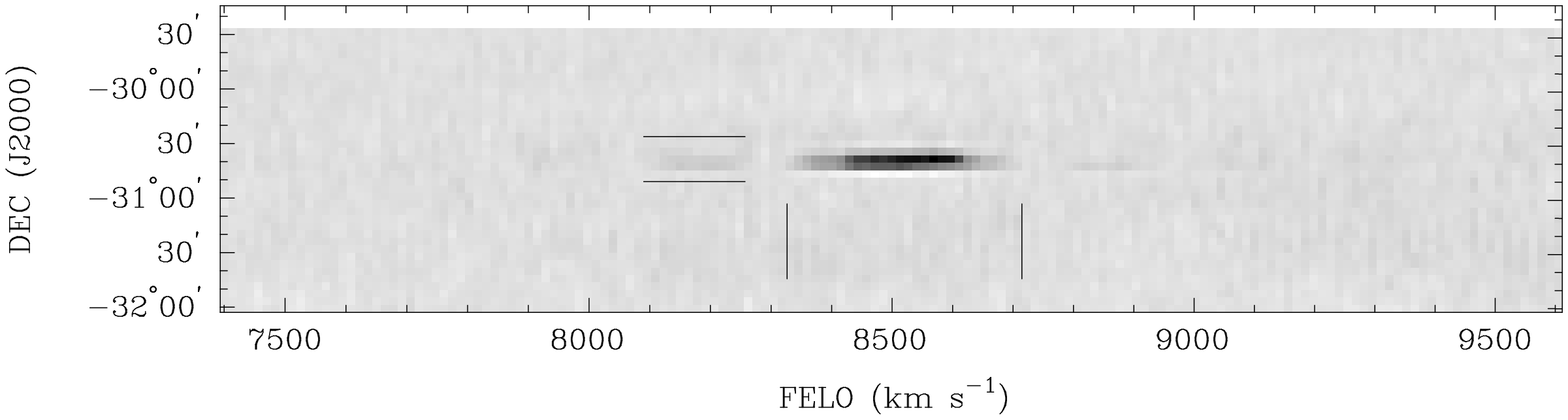}
     \end{center}
     \caption{Example spectra and position-velocity maps of GPS L3 beacon}
     \label{fig:eg_int1}
     \end{figure}
     
     \begin{figure}
     \begin{center}
     \includegraphics[trim=3.0cm 0cm 1cm 0cm,height=7.5cm,keepaspectratio=true,angle=270,clip]{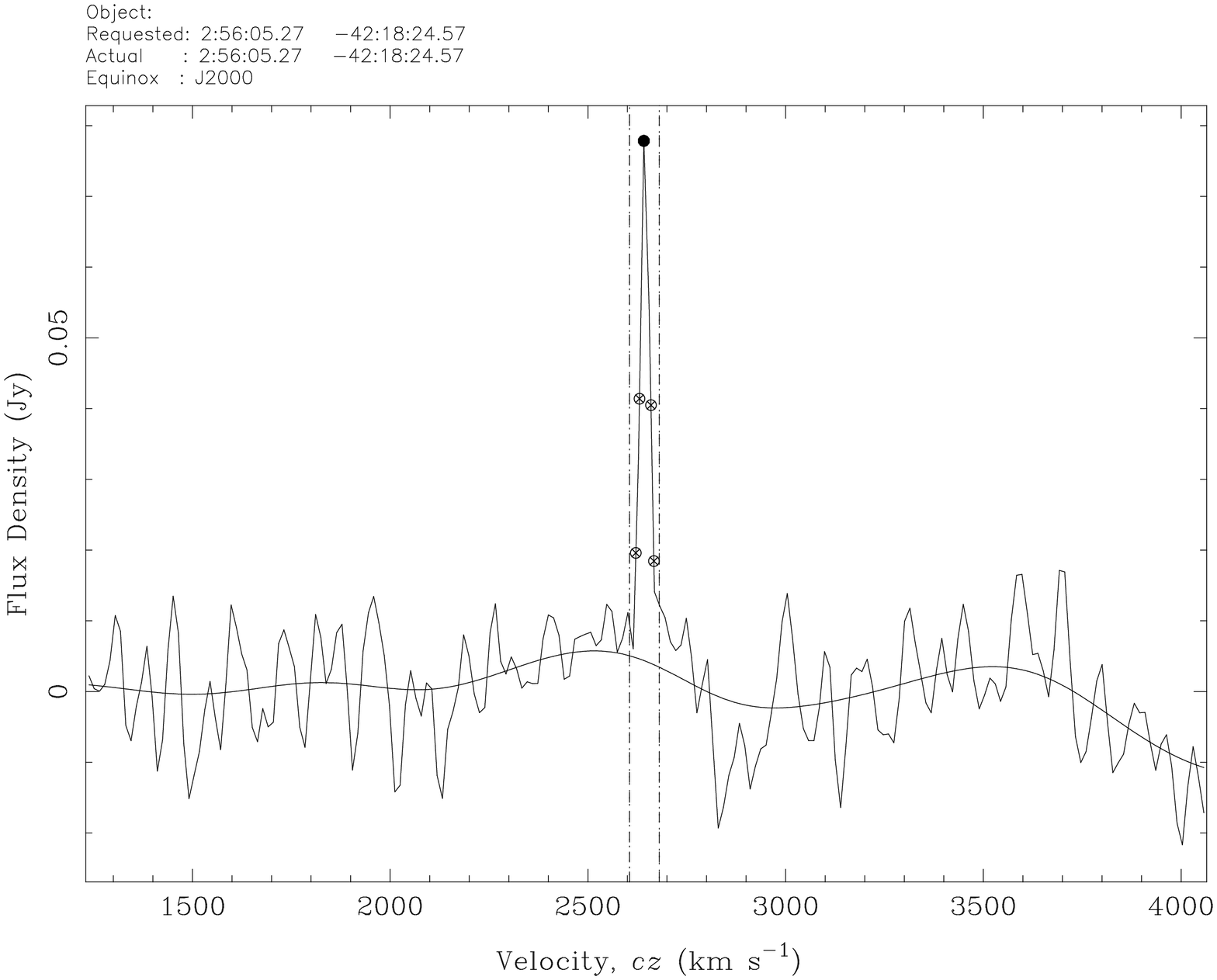}\\
     \includegraphics[trim=6.4cm 0.8cm 0cm -0.8cm,height=7.9cm,angle=270]{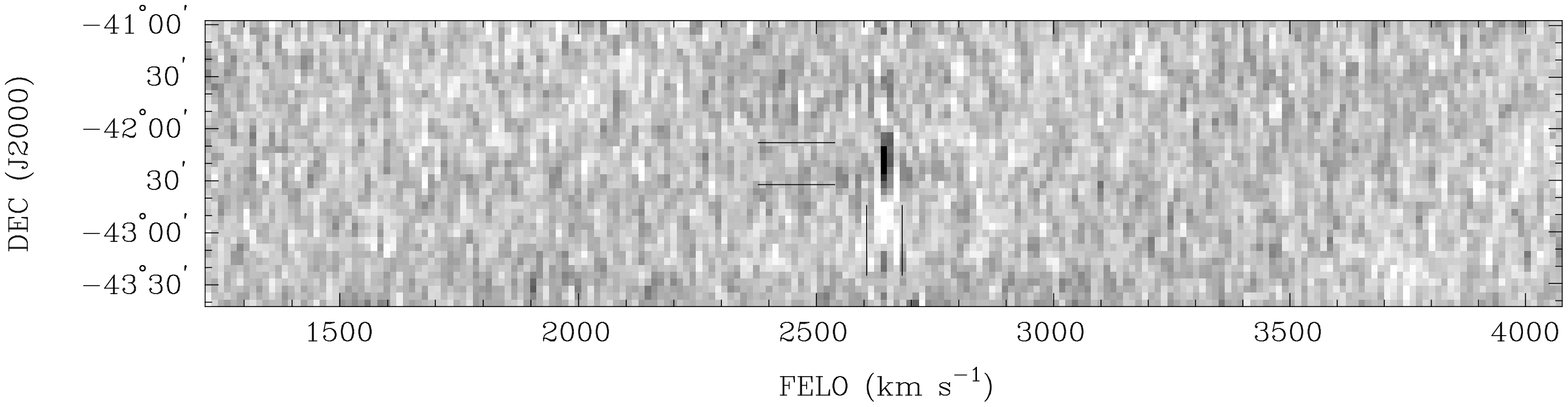}
     \end{center}
     \caption{Example spectra and position-velocity maps of narrow-line interference}
     \label{fig:eg_int2}
     \end{figure}

To minimize the potential effects of interference, scans in \hipass
were taken in 5 separate sets, each 35 arcminutes apart, at well
separated times.  Nevertheless, some narrow-band interference remains
in the data, the most prominent of which is a line at 1408 MHz
(corresponding to $\sim 2640$~\kms) caused by the 11th harmonic of the
128 MHz multibeam correlator sampler clock \citep{barnes2001}, and
another line at 1400 MHz (corresponding to $\sim 4370$~\kms) again
caused by local interference.  The presence of these, along with a
number of other narrow interference lines, is clearly evident in the
distribution of \hicat candidates following the first two checks of
the cataloguing process (top panel, Figure~\ref{fig:interference}).
Also apparent in the velocity distribution are detections associated
with the GPS L3 beacon at 1381 MHz ($\sim 8550$~\kms).  These
detections cover a range of velocities due to intrinsic signal
spreading of the GPS beacon.  The presence of these interference
detections following the first two verification checks is consistent
with the purely spectral nature of these checks and their deliberately
conservative approach to candidate rejection.  However, the
examination of position-velocity maps in the third stage of checking
enables the removal of these false detections as shown by the final
velocity distribution of the catalogue in Figure~\ref{fig:biv_obs3}.
The time varying nature of the GPS signal, combined with the HIPASS
observation and data processing methods, results in a signal that is
dominated by only a few on-sky pixels, much narrower than the signal
of an unresolved source (see Figure~\ref{fig:eg_int1}).  Similarly,
narrow-line RFI can be distinguished by the narrow-line signature
spread across large spatial regions (Figure~\ref{fig:eg_int2}).

The lack of features in the final velocity distribution at RFI
frequencies shows that \hicat completeness and reliability are not
significantly degraded at these locations.  This point is discussed
further in Section 3.5 of Paper II.

     \begin{table}
     \caption{Hydrogen $\Delta$n=1 to $\Delta$n=5 recombination
     lines in the range of \hipass observations \citep{lilley1968}
     and their apparent \hi velocities.}
     \begin{center}
     \begin{tabular}{cc}
     \hline
     Transition & Apparent \hi Velocity (\kms)\\
     \hline
     H166a &-911\\
     H209b &36\\
     H239c &438\\
     H263d &1434\\
     H283e &1540\\
     H240c &4199\\
     H210b &4340\\
     H167a &4507\\
     H284e &4718\\
     H264d &4857\\
     H285e &7918\\
     H241c &7991\\
     H265d &8306\\
     H211b &8685\\
     H168a &9990\\
     H286e &11140\\
     H266d &11781\\
     H242c &11815\\
     \hline
     \hline
     \end{tabular}
     \label{tab:recomb}
     \end{center}
     \end{table}

\subsection{Recombination lines}
\label{sec:recombination}

Another source of potentially erroneous \hi detections in \hicat is
the recombination of warm ionized gas in the Milky Way.  The full list
of hydrogen $\Delta$n=1 to $\Delta$n=5 recombination lines appearing
in the \hipass velocity range is given in Table~\ref{tab:recomb}
\citep[taken from][]{lilley1968} and plotted against the distribution
of detections following the first two verification checks in
Figure~\ref{fig:interference}).  Given the increased line strengths of
$\Delta$n=1 transitions compared to those with higher principal
quantum number differences, the erroneous detection spikes in
Figure~\ref{fig:interference} most likely due to hydrogen
recombination are H167a and H210b.  These also coincide with known RFI
frequencies.  Once again, these lines are successfully removed in the
subsequent verification stages as demonstrated by the final source
velocity distribution.

\subsection{Narrow-Band Follow-up}
\label{sec:hp_accuracy}

As detailed in Paper II, an extensive program of Parkes narrow-band
follow-up observations has been carried out to test the reliability of
\hicat.  To additionally remove as many false detections as possible,
low peak flux sources and those flagged as uncertain were
preferentially targetted.  In total, 1082 \hicat source were confirmed
in this process and 119 removed.  Detections confirmed by narrow-band
observations have been flagged as such in the final catalogue.

\section{Data}
\label{sec:data}

The full data of \hicat is available on-line at:

\vspace{2mm}
\begin{center}http://hipass.aus-vo.org\end{center}
\vspace{2mm}

This database is searchable in a number of ways, including by position
and velocity.  Returned parameters can be individually selected, along
with any of the image products, including detection spectra, on-sky
moment maps and position-velocity moment maps.  The format of the
returned catalogue data can also be chosen, with both HTML and plain
text available.  An extract of the \hipass Catalogue is given in
Table~6.

We encourage other researchers to make use of this database.  For
optimum utility, researchers also need to be aware of the
completeness, reliability and accuracy of the measured
parameters.  These are described in detail in Paper II.  Users are
also encouraged to be familiar with the full processing of \hipass
data \citep{barnes2001}.

\section{Basic Property Distributions}
\label{sec:properties}

     \begin{figure*}
     \begin{center}
     \includegraphics[trim=0cm 0cm 0cm 0cm,width=17cm,keepaspectratio=true,clip]{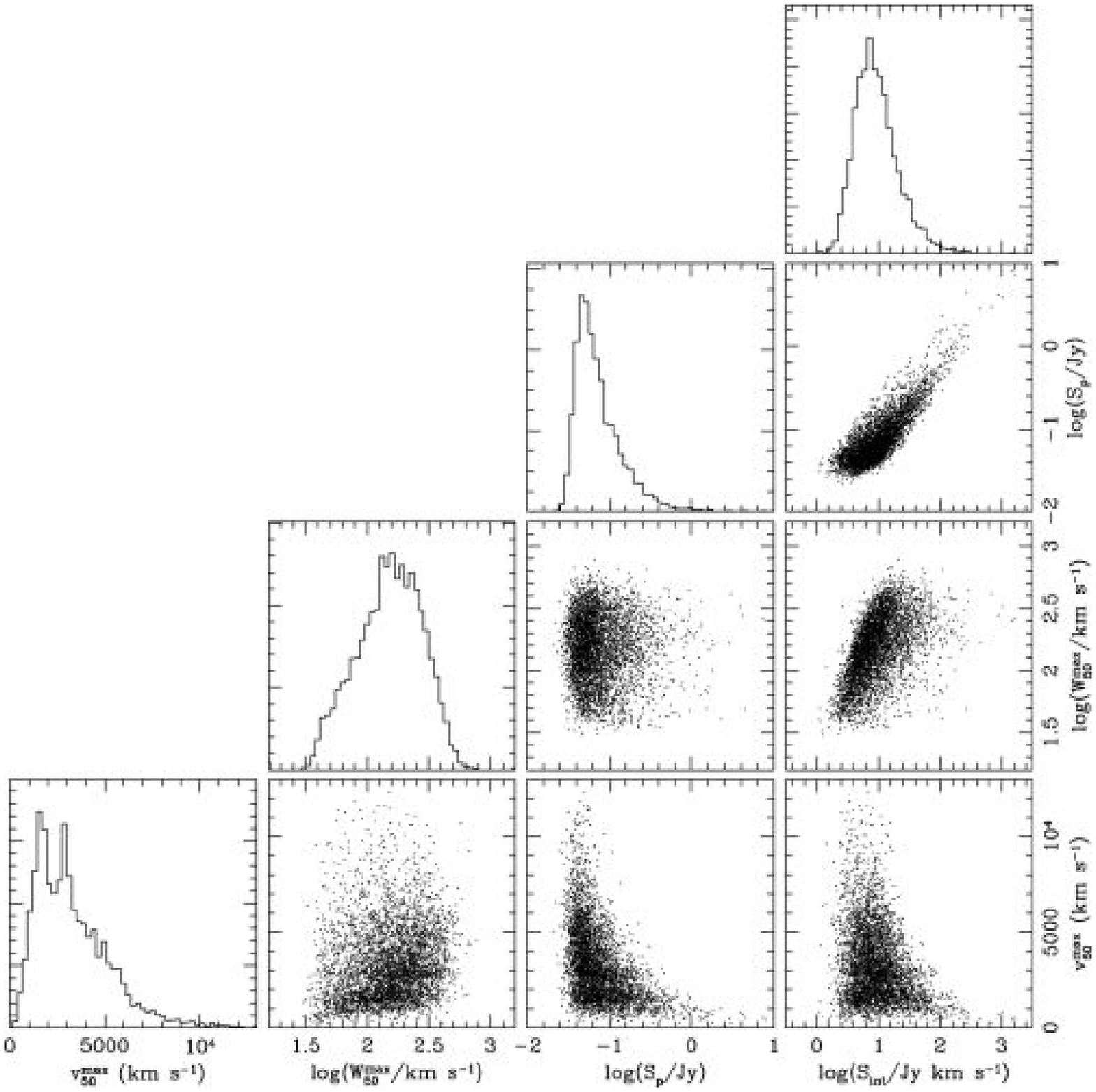}
     \end{center}
     \caption{Log-log bivariate distributions of observed parameters:
       velocity ($v_{\rm 50}^{\rm max}$), velocity width ($W_{\rm 50}^{\rm max}$),
       peak flux ($S_{\rm p}$) and integrated flux
       ($S_{\rm int}$).  Single parameter distributions are also given along
       the diagonal.}
     \label{fig:biv_obs3}
     \end{figure*}

To show the basic properties of the \hipass galaxies,
Figure~\ref{fig:biv_obs3} plots the bivariate distributions of
velocity, velocity width, peak flux and integrated flux.  Plotted
along the diagonal are the one-dimensional histograms of each
quantity.

The large-scale structure of the local universe is clearly visible in
the velocity distribution, with two prominent over-densities appearing
at $\sim$1,600 and $\sim$2,800~\kms.  This diagram also shows the
noise-limited nature of the survey, with no sharp cutoff in galaxy
numbers observed at the highest observed velocities.  From the peak
flux distribution, it is clear that most galaxies are close to the
detection limit of the survey, with the distribution peaking at
$\sim$50 mJy.

The sharp decline at the upper end of the galaxy mass function can be
seen by examining the peak flux-velocity and integrated flux-velocity
distributions.  This effective upper limit to \hi mass results in
galaxies at high distances all having low peak and integrated fluxes.
The steepness of the decline is apparent in the relatively sharp
boundary between the flux-velocity regions in which galaxies are, and
are not, observed.  The low space density of the most massive galaxies
is also visible in the width-velocity distribution, with
small-velocity-width galaxies preferentially found nearby due to the
small volumes surveyed.

     \begin{sidewaystable*}
     \footnotesize
     Table~6. Extract of the \hipass\, Catalogue.  The catalogue is presented in its entirety in the electronic version of this journal.  Parameter descriptions are given in Table~4.
     \label{tab1} 
     \begin{tabular}{lllllllllllllllllllllll}
     \hline  
     Name                    & RA                      & Dec                            & $ v_\text{50}^\text{max}$ & $ v_\text{50}^\text{min}$ & $ v_\text{20}^\text{max}$ & $ v_\text{20}^\text{min}$ & $   v_\text{mom}$        & $    v_\text{Sp}$        & $   v_\text{gsr}$        & $    v_\text{lg}$ & $   v_\text{cmb}$        & $v_\text{lo}$            & $v_\text{hi}$        \\ 
                             & (hrs)                   & (deg)                          & (\kms)                        & (\kms)                        & (\kms)                        & (\kms)                        & (\kms)                      & (\kms)                      & (\kms)                      & (\kms)               & (\kms)                      & (\kms)                      & (\kms)                  \\ 
                             & $v_\text{speclo}$        & $v_\text{spechi}$      & $v_\text{mask}$     & $ W_\text{50}^\text{max}$ & $ W_\text{50}^\text{min}$ & $ W_\text{20}^\text{max}$ & $ W_\text{20}^\text{min}$ & $     S_\text{p}$        & $   S_\text{int}$        &            RMS          &  RMS$_\text{clip}$       &  RMS$_\text{cube}$       &          cube       \\ 
                             & (\kms)                      & (\kms)                    & (\kms)                 & (\kms)                        & (\kms)                        & (\kms)                        & (\kms)                        & (Jy)                         & (Jy \kms)                   & (Jy)                    &  (Jy)                        &  (Jy)                        &                     \\ 
                             &       $\sigma$          &       box size        &  comment          &  follow-up &  confused               & extended               \\ 
                             & (\kms)                    & (arcmin)              &                   &         &                         &                        \\ 
     \hline  
     HIPASSJ0000-07 &     00:00:25.8 &     -07:49:56 &     3747.8 &     3747.8 &     3721.8 &     3721.8 &     3728.2 &     3759.8 &     3811.7 &     3870.1 &     3377.6 &     3619.0 &     3807.3 & \\ 
                      &     2210.7 &     5284.0 & 3619,3807  &     58.1 &     58.1 &     124.9 &     124.9 &     0.039 &     2.7 &     0.0065 &     0.0056 &     0.0115 &  336   \\ 
                      &   158 &     28  &  1 &     yes &     no &     no & \\ 
     HIPASSJ0000-40 &     00:00:32.3 &     -40:29:54 &     3170.8 &     3170.8 &     3169.6 &     3169.6 &     3167.9 &     3071.6 &     3138.1 &     3149.8 &     2923.4 &     2993.2 &     3342.1 & \\ 
                      &     1617.3 &     4733.8 & 2993,3342  &     239.1 &     239.1 &     258.2 &     258.2 &     0.066 &     12.0 &     0.0081 &     0.0063 &     0.0115 &  146   \\ 
                      &   158 &     28  &  1 &     yes &     no &     no & \\ 
     HIPASSJ0002-03 &     00:02:00.5 &     -03:17:01 &     6001.8 &     6001.8 &     6005.3 &     6005.3 &     6002.0 &     5966.7 &     6099.1 &     6162.8 &     5645.6 &     5917.5 &     6095.7 & \\ 
                      &     4503.7 &     7540.8 & 5918,6096  &     126.4 &     126.4 &     152.5 &     152.5 &     0.055 &     6.8 &     0.0064 &     0.0056 &     0.0115 &  337   \\ 
                      &   158 &     28  &  1 &     no &     no &     no & \\ 
     HIPASSJ0002-07 &     00:02:03.7 &     -07:37:56 &     3764.8 &     3764.8 &     3764.0 &     3764.0 &     3765.2 &     3773.3 &     3848.5 &     3907.1 &     3414.8 &     3714.6 &     3803.4 & \\ 
                      &     2307.5 &     5240.8 & 3715,3803  &     44.7 &     44.7 &     65.8 &     65.8 &     0.048 &     2.2 &     0.0073 &     0.0069 &     0.0122 &  286   \\ 
                      &   158 &     28  &  1 &     yes &     no &     no & \\ 
     HIPASSJ0002-15 &     00:02:29.5 &     -15:58:25 &     3416.2 &     3416.2 &     3420.8 &     3420.8 &     3422.6 &     3435.6 &     3478.1 &     3526.0 &     3089.5 &     3327.8 &     3532.2 & \\ 
                      &     1878.9 &     4971.3 & 3328,3532  &     93.6 &     93.6 &     147.1 &     147.1 &     0.039 &     3.7 &     0.0082 &     0.0069 &     0.0115 &  237   \\ 
                      &   158 &     28  &  1 &     yes &     no &     no & \\ 
     HIPASSJ0002-52 &     00:02:29.7 &     -52:47:15 &     1500.2 &     1500.2 &     1498.0 &     1498.0 &     1499.7 &     1504.6 &     1427.3 &     1419.4 &     1318.3 &     1416.3 &     1578.0 & \\ 
                      &     200.1 &     3030.3 & 1416,1578  &     111.4 &     111.4 &     138.1 &     138.1 &     0.051 &     5.6 &     0.0064 &     0.0055 &     0.0115 &  107   \\ 
                      &   158 &     28  &  1 &     yes &     no &     no & \\ 
     HIPASSJ0002-80 &     00:02:57.4 &     -80:20:51 &     1958.3 &     1958.3 &     1961.3 &     1961.3 &     1959.2 &     1958.3 &     1807.4 &     1758.6 &     1943.4 &     1860.9 &     2060.4 & \\ 
                      &     493.7 &     3424.0 & 1861,2060  &     98.7 &     98.7 &     132.8 &     132.8 &     0.242 &     23.4 &     0.0090 &     0.0070 &     0.0122 &  9   \\ 
                      &   158 &     28  &  1 &     no &     no &     no & \\ 
     HIPASSJ0004-01 &     00:04:46.5 &     -01:35:50 &     7194.1 &     7194.1 &     7186.9 &     7186.9 &     7186.7 &     7109.8 &     7287.7 &     7353.2 &     6829.6 &     7023.7 &     7386.3 & \\ 
                      &     5608.2 &     8795.5 & 7024,7386  &     221.4 &     221.4 &     309.4 &     309.4 &     0.046 &     9.4 &     0.0071 &     0.0067 &     0.0115 &  337   \\ 
                      &   158 &     28  &  1 &     no &     no &     no & \\ 
     HIPASSJ0005-07 &     00:05:12.3 &     -07:03:43 &     3810.0 &     3932.9 &     3816.1 &     3816.1 &     3817.7 &     3949.3 &     3901.3 &     3960.5 &     3467.6 &     3627.7 &     4045.7 & \\ 
                      &     2204.2 &     5432.7 & 3628,4046  &     314.8 &     69.1 &     346.6 &     346.6 &     0.086 &     18.0 &     0.0072 &     0.0060 &     0.0122 &  286   \\ 
                      &   158 &     28  &  1 &     yes &     no &     no & \\ 
     HIPASSJ0005-11 &     00:05:15.6 &     -11:29:37 &     6762.0 &     6762.0 &     6781.3 &     6756.0 &     6768.7 &     6750.8 &     6837.9 &     6891.6 &     6426.5 &     6652.8 &     6911.0 & \\ 
                      &     5239.8 &     8263.9 & 6653,6911  &     132.7 &     132.7 &     229.5 &     179.0 &     0.039 &     5.2 &     0.0072 &     0.0065 &     0.0122 &  286   \\ 
                      &   158 &     28  &  1 &     yes &     no &     no & \\ 
     \hline  
     \normalsize
     \end{tabular}
     \end{sidewaystable*}

     \begin{figure*}
     \begin{center}
     \includegraphics[trim=0.5cm 0.7cm 0cm 0cm,width=16.5cm,keepaspectratio=true,clip]{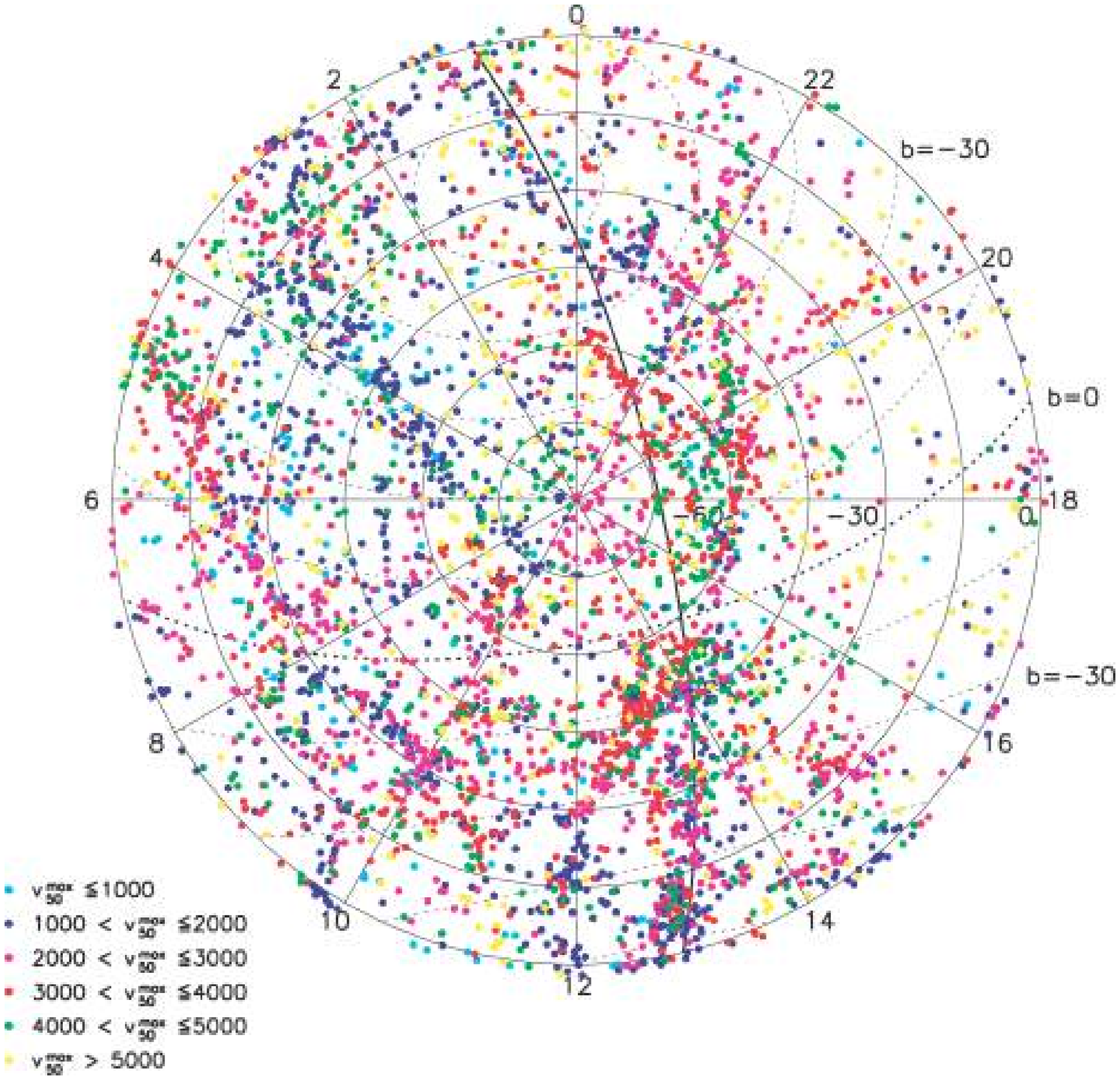}
     \end{center}
     \caption{Polar projection of \hicat detections. Objects are
       colour-coded by velocity as noted in the key.  Coordinate systems
       are marked as follows: light solid lines - equatorial coordinates; dotted
       lines - galactic coordinates; dark solid line - super-galactic plane.}
     \label{fig:skyplot_polar}
     \end{figure*}
     
     \begin{figure*}
     \begin{center}
     \includegraphics[trim=0.5cm 0.97cm 0cm 0cm,width=16.5cm,keepaspectratio=true,clip]{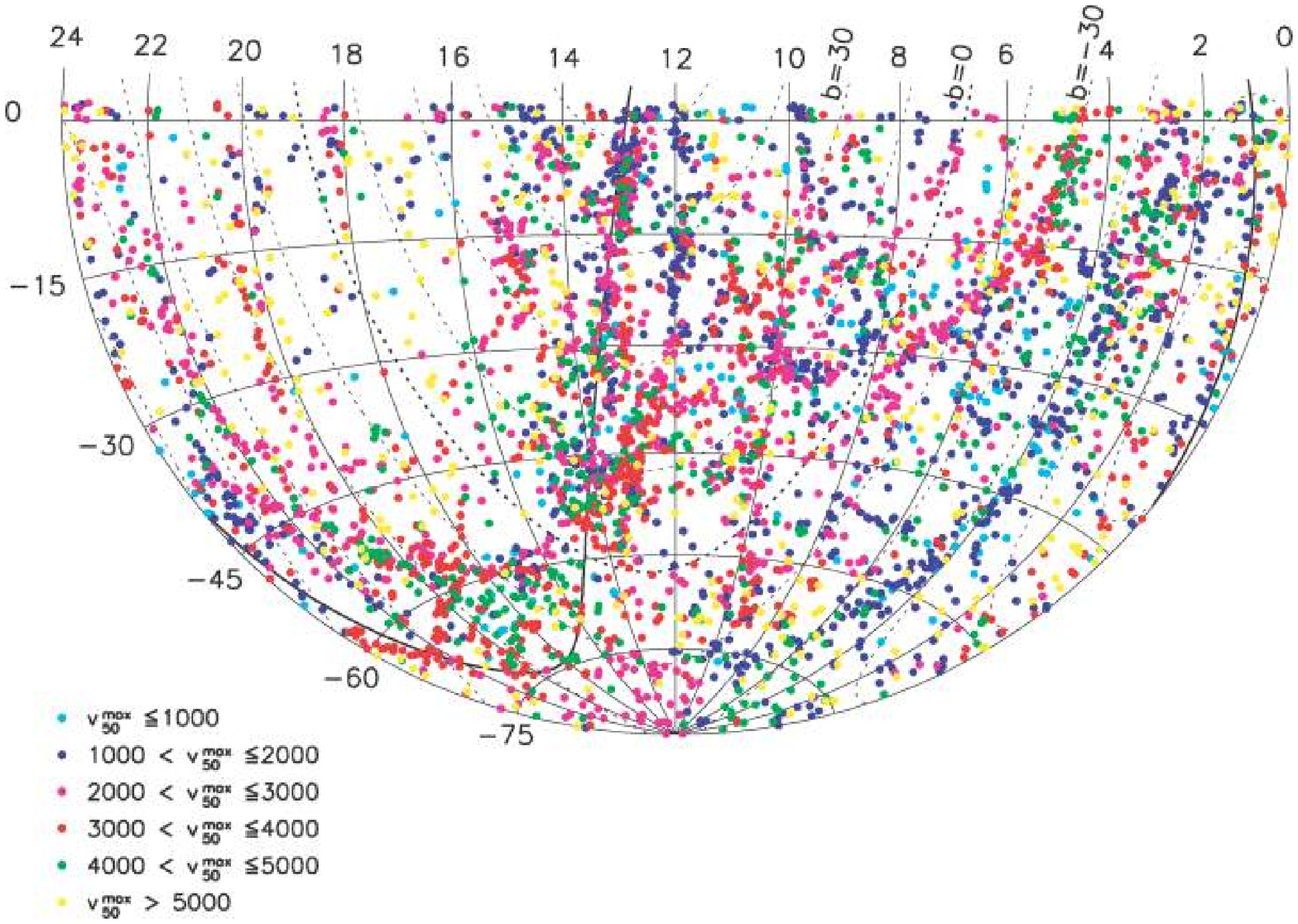}
     \end{center}
     \caption{Aitoff projection of \hicat detections. Objects are
       colour-coded by velocity as noted in the key.  Coordinate systems
       are marked as follows: light solid lines - equatorial coordinates;
       dotted lines - galactic coordinates; dark solid line - super-galactic
       plane.}
     \label{fig:skyplot_aitoff}
     \end{figure*}
    
     \begin{figure*}
     \begin{center}
     \vspace{2cm}
     \includegraphics[trim=0.15cm 0.2cm 0cm 1.5cm,width=17cm,keepaspectratio=true]{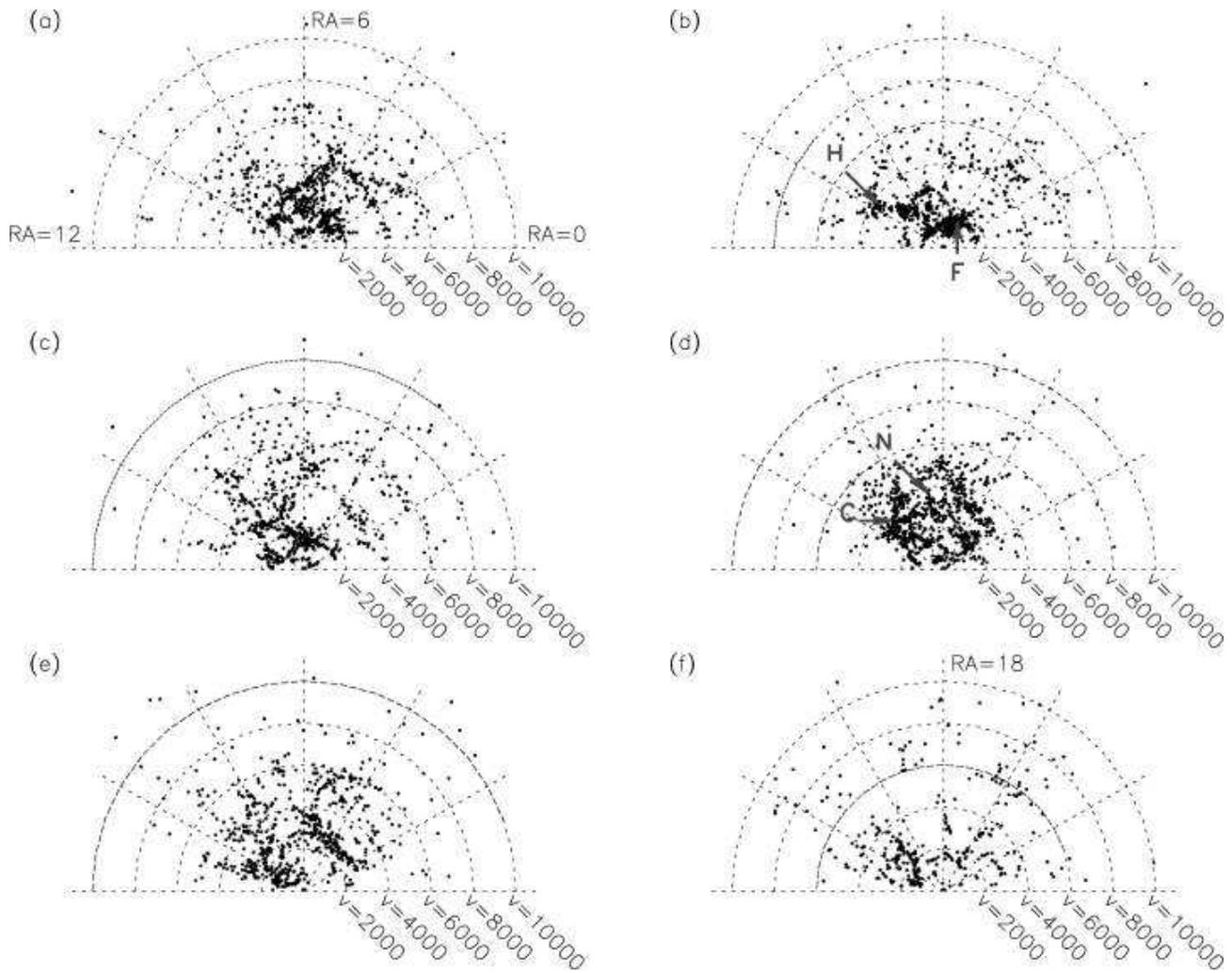}
     \end{center}
     \caption{Velocity distribution of \hicat sources divided into
     segments all hinged on the equatorial plane at the 0-12 hour line
     starting with segment (a) with one side lying in that plane
     centred at RA=6hr, (b) above that and so on with (f) again with
     one side on the plane, but centred at RA=18hr.  The location of
     these wedges in the southern sky is shown in Figure~\ref{fig:wedges}.}
     \label{fig:velplot1}
     \end{figure*}
    
     \begin{figure*}
     \begin{center}
     \vspace{2cm}
     \includegraphics[trim=0.15cm 0.2cm 0cm 1.5cm,width=18cm,keepaspectratio=true]{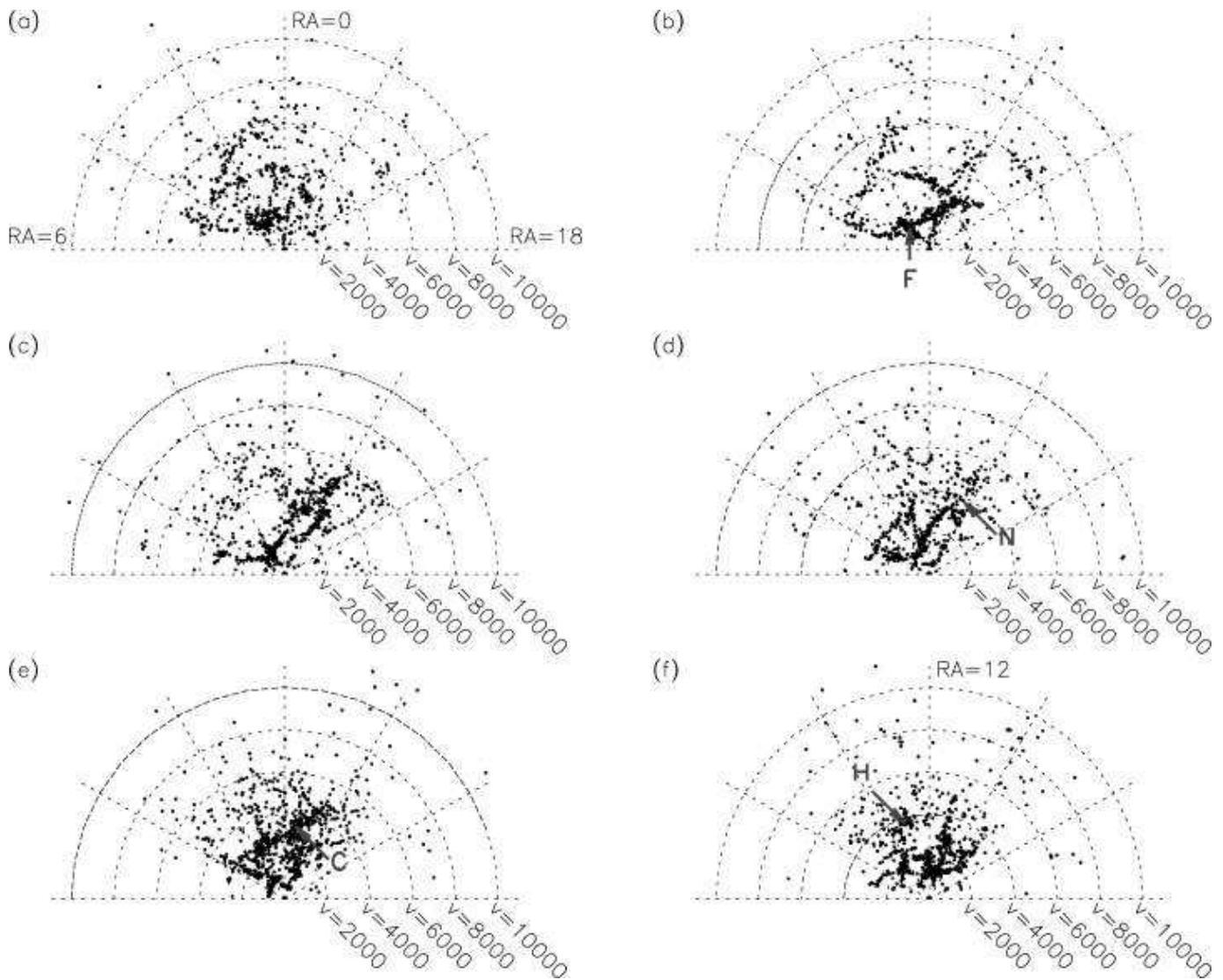}
     \end{center}
     \caption{Velocity distribution of \hicat sources divided into
     segments all hinged on the equatorial plane at the 6-18 hour line
     starting with segment (a) with one side lying in that plane
     centred at RA=0hr, (b) above that and so on with (f) again with
     one side in the plane, but centred at RA=12hr.}
     \label{fig:velplot0}
     \end{figure*}

     \begin{figure*}
     \begin{center}
     \includegraphics[trim=6cm 20cm 3cm 0cm,width=7cm,keepaspectratio=true]{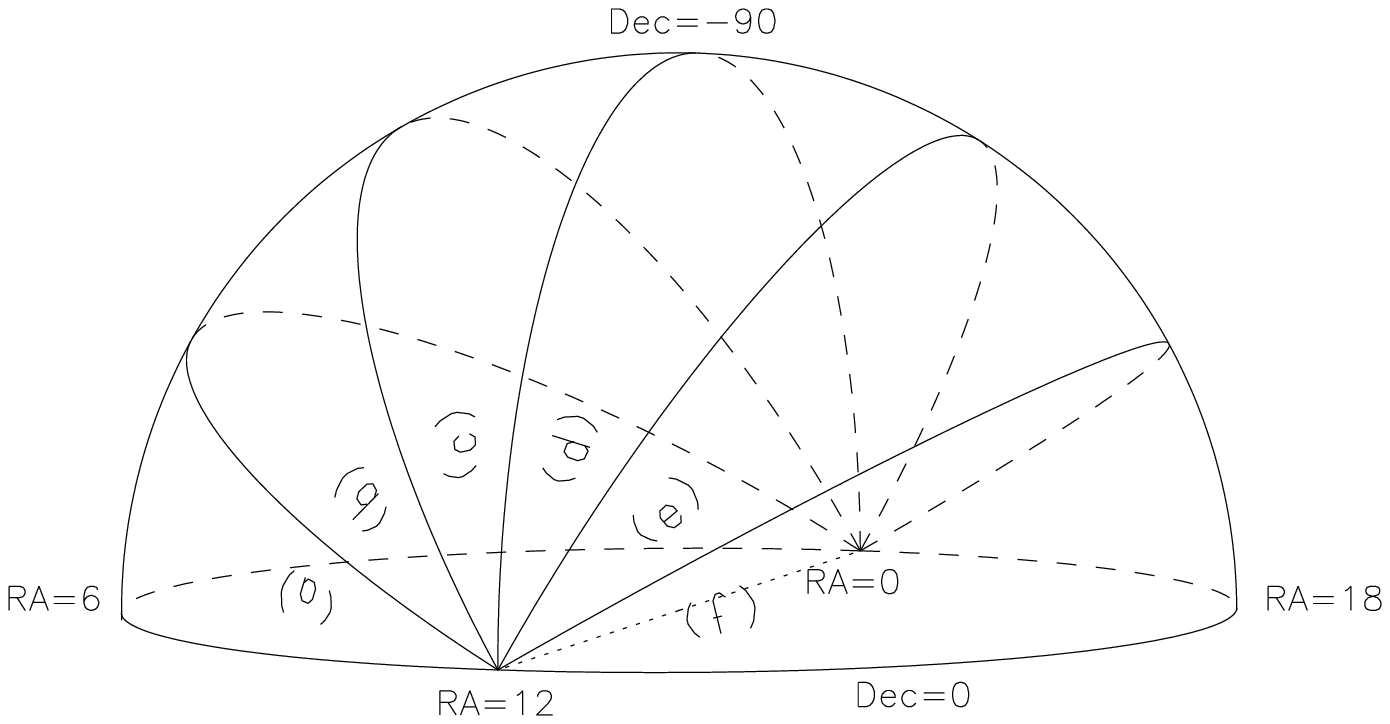}
     \end{center}
     \caption{Location in the southern sky of the velocity
     distribution wedges shown in Figure~\ref{fig:velplot1}.}
     \label{fig:wedges}
     \end{figure*}

\section{Large Scale Structure}
\label{sec:lss}

Figures~\ref{fig:skyplot_polar}~and~\ref{fig:skyplot_aitoff} plot the
equatorial coordinate polar and Aitoff projections of the full on-sky
distribution of \hicat sources respectively.  Points are shaded
according to the observed recessional velocities.  From these plots,
the inhomogeneous distribution of HI sources in the local universe is
clearly apparent.  Most obvious are the Super-Galactic plane (marked
with a dark solid line), the Fornax wall (comprising the
over-densities around Dorado, Fornax and Eridanus; passes through
RA=3.5 hrs, Dec=$-30\degr$), and a third filament extending from near
the Super-Galactic plane, through Antlia, Puppis and Lepus (passes
through RA=8 hrs, Dec=$-30\degr$).  Also notable are significantly
under-dense regions, the most prominent of which is the Local Void
(around RA=18 hrs, Dec=$-30\degr$).  The location of large-scale
structures is important when determining the representative nature of
galaxy surveys given their search region.

\hicat also maps structures behind the Milky Way, with relatively
little obscuration apparent in the observed distribution.  Many of
these structures and filaments have been previously identified
\citep{kraan2000,henning1998}, however their uninterrupted path
through the Zone of Avoidance is traced out here for the first time in
a single survey.

In Figures~\ref{fig:velplot1}~and~\ref{fig:velplot0}, the velocity
distribution of \hicat galaxies is plotted in two different projection
sets.  In Figure~\ref{fig:velplot1}, the southern sky hemisphere is
divided like segments of an orange into six wedges, each with a common
base axis along the line RA=0-12 in the equatorial plane.  The
location of these wedges with respect to the southern celestial
hemisphere is shown in Figure~\ref{fig:wedges}.  The shape of these
regions results in volume increasing in the vertical rather than
radial direction.  Thus, more galaxies are present in the central
horizontal regions when compared to similar velocity regions further
out.  Figure~\ref{fig:velplot0} also divides the southern hemisphere
into six wedges, this time using a common base axis along RA=6-18 in
the equatorial plane.

These figures again show the Super-Galactic plane and the Local Void.
Figure~\ref{fig:velplot1}(d) contains the bulk of the Super-Galactic
plane in face-on projection and Figures~\ref{fig:velplot0}(c)-(e)
show the plane in edge-on cross sections.  From this second set of diagrams
it is also clear that local regions of the plane are in fact comprised
of two main parallel components.  The lower right regions of these
diagrams show the significantly under-dense region of the Local
Void.  Figure~\ref{fig:velplot1}(f) most clearly shows this in the
orthogonal projection in the centre of the diagram.  Also marked in
the velocity diagrams are the locations of four major nearby clusters
(F=Fornax, C=Centaurus, H=Hydra and N=Norma), their locations
coinciding with over-dense regions in the \hicat data.

\section{Summary}

\hipass is the first blind \hi survey of the entire southern sky.
Using the Multibeam receiver on the Parkes Telescope, this survey
covers $\delta < +25\degr$ with a velocity range $-$1,280 to
12,700~\kms.  Mean noise for the survey is 13 mJy beam$^{-1}$.  Taking
data from the southern region of this survey $\delta < +2\degr$, we
have compiled a database of 4,315 sources selected purely on their \hi
content.  Candidates are found using automatic finder algorithms,
before being subjected to an extensive process of manual checking and
verification.  Issues examined in compilation of the catalogue include
RFI and recombination lines, along with additional processes for
confused and extended sources.  A large program of follow-up
observations is also used to confirm catalogue sources.  Completeness,
reliability and parameter accuracy of \hicat sources is addressed in
Paper II.  Optical counterparts to \hipass galaxies are examined in
Paper III.

The large-scale distribution of sources in the catalogue covers a
number of major local structures, including the Super-Galactic plane
along with filaments passing through the Fornax and Puppis regions.
Also notable is the probing of structure in the Zone of Avoidance, a
region significantly obscured in optical bands.  The survey also
surveys a number of under-dense local regions including the Local Void.

This database offers a unique opportunity to study a number of
important astrophysical issues.  Future papers will address topics
such the cosmic mass density of cool baryons, the properties of \hi
selected galaxies at other wavelengths, the comparative clustering of
\hi-rich galaxies, the Tully-Fisher relation, the role of hydrogen in
galaxy group evolution and the study of \hi-rich galaxy environments.

\section{Acknowledgments}

The Multibeam system was funded by the Australia Telescope National
Facility (ATNF) and an Australian Research Council grant. The
collaborating institutions are the Universities of Melbourne, Western
Sydney, Sydney and Cardiff, Research School of Astronomy and
Astrophysics at Australian National University, Jodrell Bank
Observatory and the ATNF. The Multibeam receiver and correlator was
designed and built by the ATNF with assistance from the Australian
Commonwealth Scientific and Industrial Research Organisation Division
of Telecommunications and Industrial Physics. The low noise amplifiers
were provided by Jodrell Bank Observatory through a grant from the UK
Particle Physics and Astronomy Research Council. The Multibeam Survey
Working Group is acknowledged for its role in planning and executing
the HIPASS project.  This work makes use of the AIPS++, \miriad and
{\sc Karma} software packages.  We would also like to acknowledge the
assistance of Richard Gooch, Mark Calabretta, Warwick Wilson and the
ATNF engineering development group.  Finally, we thank Brett Beeson
and the Australian Virtual Observatory for development of the {\sc
SkyCat} service which hosts the \hicat data.


\bibliographystyle{mn2e}
\bibliography{mn-jour,all}

\label{lastpage}
\end{document}